\documentclass[aps, prd, reprint, superscriptaddress, amsmath, amssymb]{revtex4-2}

\usepackage{microtype}
\usepackage[normalem]{ulem} 
\usepackage{enumitem}

\usepackage{mathtools}
\usepackage{xfrac}
\usepackage{cancel}
\usepackage{braket}

\usepackage{graphicx}
\graphicspath{{figures/}}
\usepackage[caption=false]{subfig}

\usepackage[usenames, dvipsnames, svgnames, table]{xcolor}
\usepackage{comment}
\usepackage{linenoAMS}

\usepackage[
    range-phrase=--,
    range-units=single,
    retain-explicit-plus=true,
    retain-unity-mantissa=false
]{siunitx}

\usepackage[colorlinks=true, citecolor=RoyalBlue, linkcolor=BrickRed, urlcolor=ForestGreen]{hyperref}
\usepackage[all]{hypcap}

\usepackage[capitalise]{cleveref}

\DeclareMathOperator{\sinc}{sinc}
\definecolor{spring}{rgb}{0.7,0.9,0.7}
\definecolor{brick}{rgb}{0.7,0.2,0.1}
\definecolor{redHL}{rgb}{1.0,0.7,0.7}
\definecolor{blueHL}{rgb}{0.7,0.7,1.0}
\definecolor{ElectricBlue}{rgb}{0.49, 0.976, 1.0}
\definecolor{blueT}{rgb}{0.039, 0.729, 0.71}
\definecolor{DarkGreen}{rgb}{0, 0, 0}
\definecolor{blue}{rgb}{0.066, 0.310, 0.984}
\definecolor{blueT}{rgb}{0.039, 0.729, 0.71}
\definecolor{black}{rgb}{0, 0, 0}
\definecolor{violet}{rgb}{0.749, 0.467, 0.965}

\usepackage[capitalise]{cleveref}
\usepackage{commands}

\begin{document}

\title{Gravitational-wave astronomy with a space-based optical clock network}
\author{Dhruva Ganapathy}%
\email{dhruva@berkeley.edu}
\author{Divya Singh}%
\author{Sammith Singamsetty}%
\author{Shimon Kolkowitz}%

\affiliation{Department of Physics, University of California, Berkeley, CA 94720, USA
}%

\overfullrule 0pt 
\parskip0pt
\hyphenpenalty9999

\date{\today}
\begin{abstract}
Since the first detection of a merging binary black hole system a decade ago, gravitational-wave astronomy has emerged as a powerful tool for astrophysics. Future space-based observatories, such as the Laser Interferometer Space Antenna (LISA), will unlock the millihertz (mHz) band, which remains entirely inaccessible to ground-based detectors due to terrestrial noise. In parallel, proposed atom-based gravitational-wave detectors, specifically those based on space-based optical clocks and atom interferometers, offer capabilities that are unique and complementary to traditional optical interferometers. Their highly tunable character enables sensitive measurements across a broad frequency band extending from the mHz up to and possibly even above the Hz regime. In this work, we investigate the use of one-way Doppler tracking in space-based atomic clock networks operating in concert with detectors like LISA. We analyze the dominant noise sources and develop dedicated measurement protocols, which operate in a targeted mode, using an externally supplied signal model to track the response coherently as the source chirps. Performing preliminary parameter estimation on simulated signals, we demonstrate how these detectors could be used to extract critical astrophysical information about binary gravitational-wave sources. We further show that the same instrument retains sensitivity in a search mode, where no prior on the source is assumed.
\end{abstract}

\maketitle



\section{Introduction}
The current era of gravitational-wave astronomy, heralded by the LIGO interferometers' first direct detections of merging black-holes in Ref.~\cite{Abbott_2016}, has reshaped our understanding of the universe. With the international gravitational-wave network having recently concluded its fourth observing run (O4), the number of detected gravitational wave events now numbers nearly 400 \cite{GWTC3,GWTC4,GraceDB_O4}. This wealth of data has provided profound insights into black hole population demographics \cite{Populations}, the extreme physics of neutron star collisions \cite{170817Tidal,170817EOS}, and the nucleosynthesis of heavy elements \cite{Kaisen2017}, and has also hinted at the existence of elusive intermediate-mass black holes \cite{IMBH}. Crucially, these observations also continue to inform our understanding of the nature of gravity \cite{TestsRelativity} and our models of cosmic expansion \cite{CosmicExpansionHistory}.

While future observing runs and proposals for third-generation terrestrial detectors, such as Cosmic Explorer \cite{evanshorizon} and the Einstein Telescope \cite{Punturo_2010_ET}, hold immense promise for the advancement of gravitational-wave astronomy, the detection band of ground-based interferometers is fundamentally limited by seismic and technical noise at low frequencies \cite{Aasi2015,Harms2019}. To overcome this limitation, space-based observatories such as the Laser Interferometer Space Antenna (LISA) are being developed to access the astrophysically rich millihertz (mHz) frequency band. This regime holds the key to understanding supermassive black holes, the galactic structure of the Milky Way, and the stochastic gravitational-wave background \cite{LISAredbook}. LISA's  sensitivity \cite{Robson_2019}, however, drops sharply above 10 mHz, leaving the decihertz-Hz region of the gravitational band largely unexplored. This frequency band is thought to contain important information about intermediate-mass black hole (IMBH) binaries, and the early inspiral phase of binary neutron star mergers that could prove crucial in being able to locate their kilonova counterparts during EM follow ups \cite{Sedda_2020_decihertz}. To bridge this observational gap, mission concepts such as DECIGO \cite{Kawamura2011} and BBO \cite{Harry2006} have been put forward. More recently, quantum atomic sensors have been identified as promising candidates for
this intermediate band, owing to their highly tunable designs and unique frequency
responses \cite{AtomsGW,Kolkowitz16,Abe21,Vutha2015,SingleAIGW}. Proposals in this
direction share a common principle; atoms furnish a precise phase or frequency
reference for light propagating across a long baseline. Atom-interferometric schemes
read the perturbation out as a phase accumulated between spatially separated atomic wave-packets \cite{Dimopoulos2008}, with proposals for both terrestrial and
space-based instruments \cite{Abe21,AION,AEDGE,SAGE} while clock-based schemes use the internal states of confined atoms to detect frequency shifts of light transmitted along the link \cite{Kolkowitz16,Vutha2015}. In
this work we develop the latter, exploring a space-based network of one-way frequency
comparisons between optical lattice clocks on a constellation of satellites, whose
response differs qualitatively from the round-trip measurements of interferometric
detectors, as we derive in \cref{sec:detector_response}. While prior work analyzing atom-based gravitational-wave detection has mainly relied on sensitivity limits based on the quantum projection limit and Fisher information, developing and realizing these detectors in practice will require connecting theoretical bounds to real data analysis. We build on the earlier proposals by developing a thorough framework for extracting astrophysically relevant information from these networks using concrete measurement protocols. 

The detection of gravitational waves with clock measurements relies on the principle that a passing gravitational wave will modulate the arrival time of light waves across an optical link. This one-way Doppler response was first formalized by Estabrook and Wahlquist in 1975 \cite{OGderivation}, and later utilized by Hellings and Downs to describe the cross-correlation signature of a stochastic gravitational-wave background in pulsar timing arrays \cite{HellingsDown}. Building on this theoretical foundation, international pulsar timing array collaborations, including NANOGrav, have recently published the first strong evidence for a stochastic gravitational-wave background at nanohertz (nHz) frequencies \cite{Agazie2023,Antoniadis2023,Reardon2023,Xu2023}.

Our ability to make precise timing and frequency measurements fundamentally determines our sensitivity to gravitational waves. Recent experimental progress has enabled unprecedented frequency stability in optical lattice atomic clocks \cite{LudlowReview,Aeppli24,Oelker19}, specifically those utilizing the long-lived ~\singletS$-$~\tripletPzero transition in ~\Sra~\cite{Courtillot03}. We note that this ultra-narrow transition has also been used for atom-interferometry \cite{Hu2017SrIFO}. Furthermore, differential frequency comparisons utilizing a shared clock laser across multiple atomic ensembles allow us to bypass the stringent limits traditionally set by laser coherence times \cite{Zheng22,Zheng23}.

We begin in \cref{sec:diff_comp} by detailing the experimental setup and the mechanics of differential clock comparisons. In \cref{sec:detector_response}, we derive the rigorous response of a one-way optical link to passing gravitational waves. \cref{sec:detector_noise} then analyzes the dominant noise budgets, extending earlier
treatments \cite{Kolkowitz16} with a thorough accounting of residual laser noise in
realistic pulse sequences, and with the in-band stability requirements imposed by
time-varying systematic shifts of the clock transition. \cref{sec:protocols} introduces a dedicated measurement protocol for estimating the phase and amplitude of a detected gravitational-wave signal, together with the interrogation timing required to average coherently over a chirping source, the precision with which the signal model must be supplied, and a search mode requiring no prior knowledge of the source. Building upon this framework, \cref{sec:pe} describes a proposed network of clock detectors in heliocentric orbit and demonstrates how this protocol enables preliminary parameter estimation for binary sources. Finally, we highlight the network's astrophysical potential by presenting an example science case focused on the spatial localization of intermediate-mass binary black hole systems.

\section{Differential Clock Comparisons}
\label{sec:diff_comp}
\begin{figure}
    \centering
    \includegraphics[width=\linewidth]{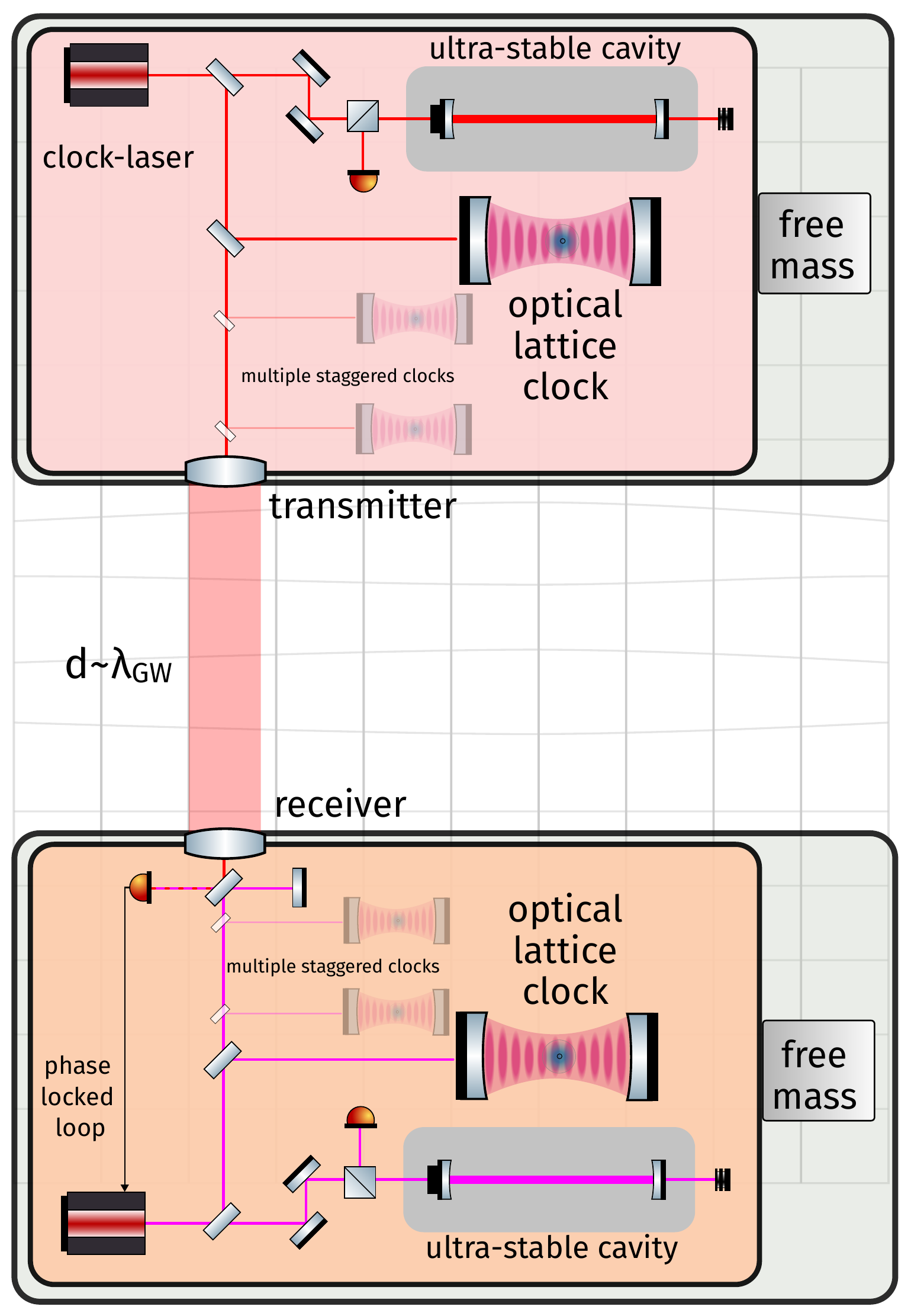}
    \caption{Differential clock comparisons across separated satellites for measuring gravitational waves. Each satellite houses a set of optical lattice clocks, kept in free fall by referencing them to a drag-free mass. The clock laser on the receiving satellite is phase-locked to the laser on the transmitting satellite for common-mode noise rejection. The lasers are additionally referenced to ultra-stable optical cavities in order to minimize excess laser noise across the optical link.}
    \label{fig:setup}

\end{figure}

In this work, we study the use of differential clock comparisons to sense a gravitational wave, learn its amplitude and phase, and localize its source. In such a setup (shown in \cref{fig:setup}), two atomic ensembles separated by a long baseline are probed with the same clock laser. This allows us to greatly exceed the interrogation times traditionally limited by finite clock laser coherence, which remains the primary constraint on the instability of state-of-the-art optical clocks \cite{LudlowReview,Oelker19}. In the context of space-based gravitational-wave detection, this is achieved by phase-locking the clock lasers of two distant satellites across the optical baseline. To minimize residual laser noise from the phase-locked loop, the clock laser in each satellite is additionally referenced to an ultra-stable optical cavity. While fully operational spaceborne optical lattice clocks remain an open developmental goal, significant milestones toward their realization have been achieved recently in a compact~\Sra~subsystem payload aboard the Chinese Space Station \cite{Guo2021,Xia2025,Yu2025}.

A metric perturbation induced by a gravitational wave propagating between the transmitter and receiver gives rise to a fractional frequency shift in the transmitted light. This shift is measured by conducting precision spectroscopy on both atomic ensembles. The clock measurements on the two ensembles are offset by the light propagation time in order to keep the laser phase noise common mode. The signature of the gravitational wave is added to the laser link during propagation. We note here that such a one-way photon link measurement has a significantly different detector response \cite{Kolkowitz16} compared to round-trip detectors like LIGO. We discuss this response in detail in \cref{sec:detector_response}.

In optical lattice clocks, atoms are tightly confined in the trapping potential of a high-intensity, magic-wavelength lattice laser. Measurements of gravitational radiation, however, require the test masses to be in free-fall. To achieve this, the lattice and clock lasers are firmly referenced to a drag-free mass. The state-of-the-art capabilities of this technology were demonstrated by the LISA Pathfinder mission, which achieved a residual acceleration noise below $2 \times 10^{-15} \text{ m s}^{-2} / \sqrt{\text{Hz}}$, exceeding the drag-free control requirements needed to detect gravitational waves in the millihertz to decihertz band \cite{LISA_Pathfinder,femtoG_pathfindfer,LISAaccelerationnoise}. In \cref{sec:detector_noise}, we analyze the effect of residual acceleration in these drag-free satellites on the clock measurements.

\begin{figure}
    \centering
    \includegraphics[width=\linewidth]{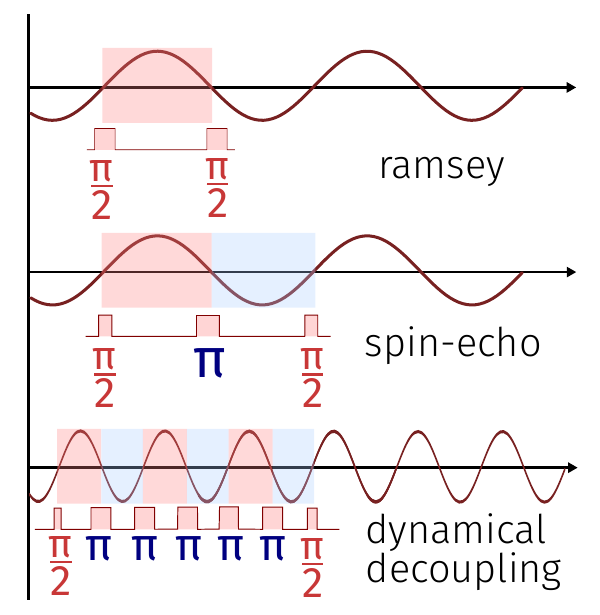}
    \caption{Interrogation sequences for sensing an oscillating signal. Each sequence begins and ends with a $\pi/2$ pulse, which respectively creates and reads out a superposition of the two clock states, while intervening $\pi$ pulses reverse the sign of the accumulated phase. Shading indicates the sensitivity function, with positive (red) and negative (blue) regions. Ramsey, spin-echo, and dynamical-decoupling sequences with $N_p$ $\pi$ pulses span half a cycle, one cycle, and $(N_p+1)/2$ cycles, respectively.}
    \label{fig:pulse_sequences}
\end{figure}

For oscillating signals such as gravitational waves, spin-echo and dynamical-decoupling pulse sequences are applied to the atoms using the local clock lasers, serving as narrow-band filters to target specific signal frequencies \cite{Kolkowitz16}. These sequences are illustrated in \cref{fig:pulse_sequences}. Each begins with the atoms prepared in one of the two clock states. A $\pi/2$ pulse drives each atom into an equal superposition of the two clock states, which subsequently accumulates phase at a rate set by the detuning between the laser and the atomic transition. A $\pi$ pulse exchanges the amplitudes of the two states and thereby reverses the sign of the phase accumulated thereafter, so that the contributions from successive half-cycles of the signal add coherently rather than averaging away; the number of $\pi$ pulses sets the number of half-cycles spanned. A final $\pi/2$ pulse converts the accumulated phase difference into a population difference, which is read out by fluorescence detection.

To enable the phase and amplitude estimation protocols described later in \cref{sec:protocols}, each satellite will need the ability to make multiple clock measurements during overlapping, staggered interrogation periods. This requirement can be satisfied in several ways. A straightforward but resource-demanding option is for each satellite to house multiple independent lattice clocks. Alternatively, the ability to independently interrogate multiple atomic ensembles within a single optical lattice would significantly ease these resource requirements \cite{Zheng24}. Continuously loaded clocks \cite{Katori_2021,Katori_CL} provide yet another alternative, wherein a continuous atomic beam could, in principle, provide an arbitrarily large number of staggered clock measurements.

At the end of a differential spectroscopy sequence, the atomic populations for the two ensembles (labeled $a$ and $b$) are given by, 

\begin{equation}
P_a = \frac{1}{2}(1 + C \cos(\phi_C)), \quad
P_b = \frac{1}{2}(1 + C \cos(\phi_C + \phi_D)),
\label{eq:diff_meas}
\end{equation}

where $\phi_C$ and $\phi_D$ correspond to the common and differential phase accumulated over the interrogation period. $C$ corresponds to the contrast of the atomic ensembles. Typically, multiple experimental shots are collected, and methods such as ellipse fitting
and maximum likelihood estimation (MLE) are used to extract the value of $\phi_D$ jointly
with the contrast $C$. With erasure conversion \cite{Niroula24,Erasure25}, coherence times, and subsequently interrogation times, can be extended to be of the order of the natural lifetime of the atomic clock transition, which is around 160 s for the ~\tripletPzero$-$~\singletS transition in \Sra~\cite{SrLifetime,Kim25_wannier_stark}. Zero dead-time operation is essential to continuously average a gravitational wave signal, which can be achieved by interleaving clocks \cite{Zero-Dead-Time2013,two_ensemble}, or by using continuously loaded clocks.

\section{Detector Response}
\label{sec:detector_response}
 \begin{figure}
    \centering
    \includegraphics[width=\linewidth]{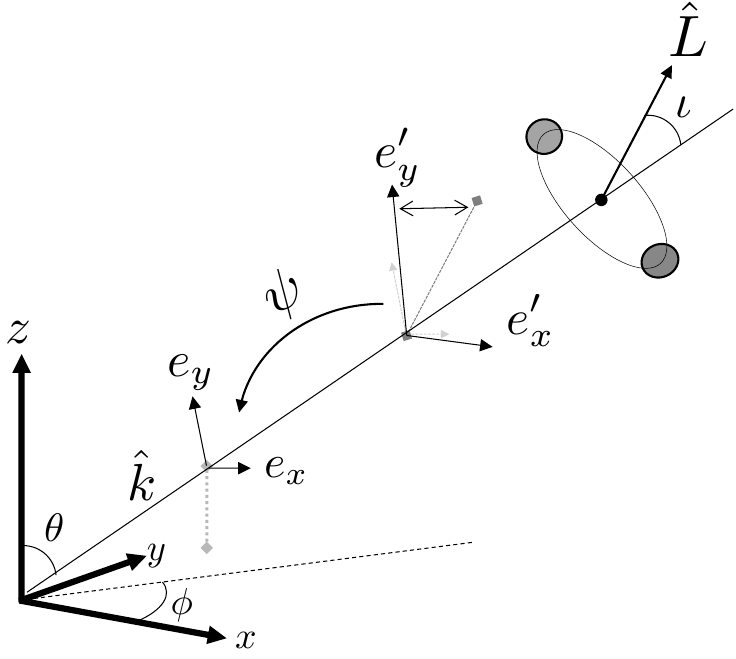}
    \caption{Coordinate system to establish the geometry of a gravitational wave source (a merging compact binary in this case). $\theta$ and $\phi$ correspond to the sky-location of the source, which fix the gravitational wave-vector $\hat{k}$, and the detector polarization basis $(e_x,e_y)$. The source's `natural' basis $(e'_x,e'_y)$ is set by the binary's angular momentum, $\hat{L}$, and is rotated from the detector's basis by a polarization angle $\psi$.}
    \label{fig:geometry}
\end{figure}

\begin{figure*}
  \centering
  \subfloat[$f$ = 1 mHz \label{fig:antenna_LF}]{
    \includegraphics[width=\textwidth]{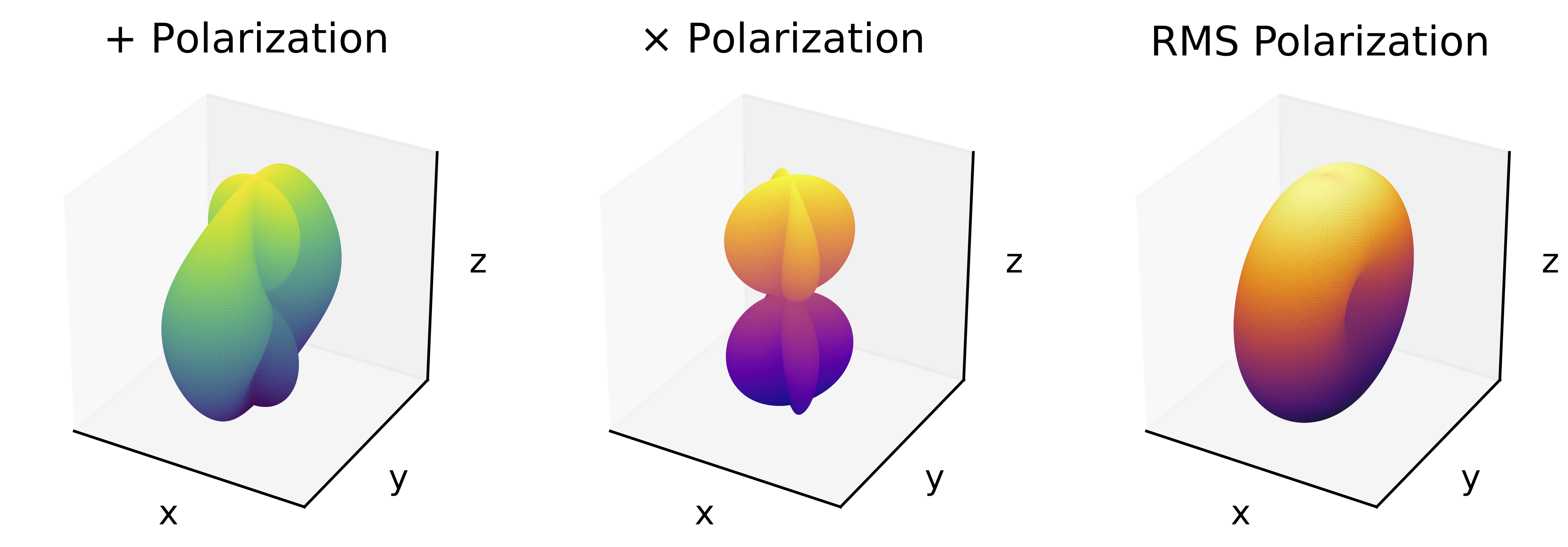}
  } \\
  \subfloat[$f$ = 100 mHz \label{fig:antenna_HF}]{
    \includegraphics[width=\textwidth]{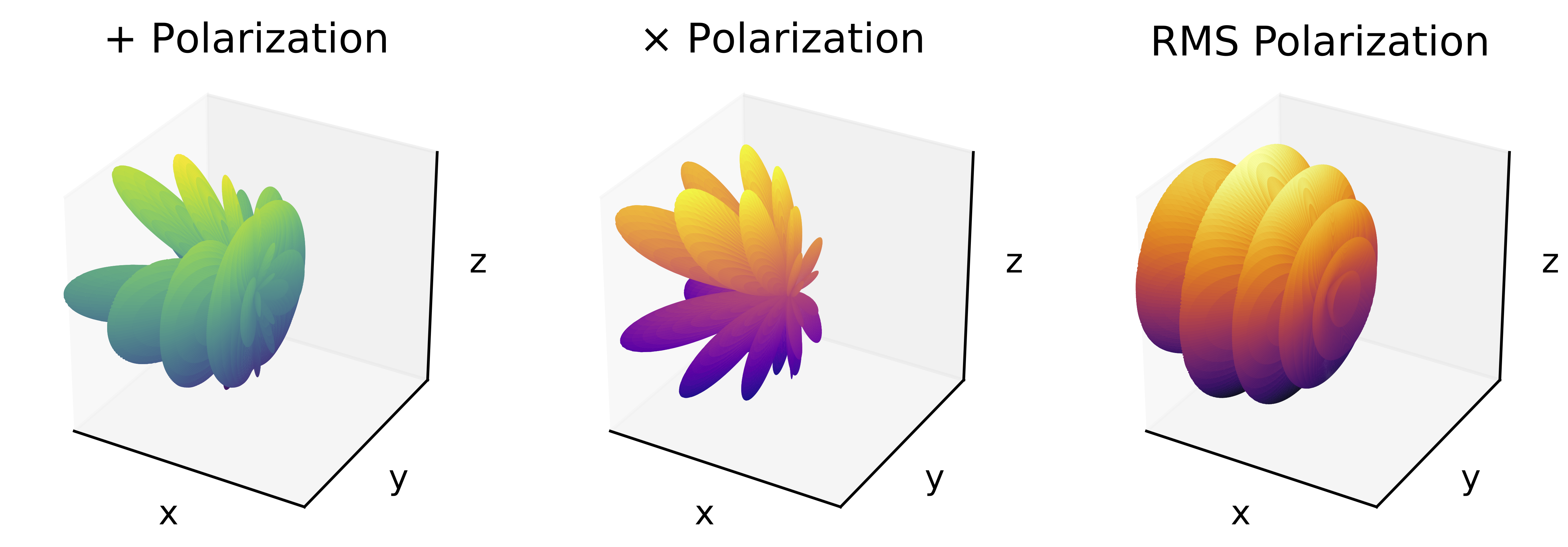}
  }
  \caption{Antenna pattern for a one-way link of length L = $1\times10^{10}$ m, calculated using \cref{eq:antenna_response}. At $f = 1$ mHz, we are in the long-wavelength limit ($fL/c\ll1$), where the transfer function is decoupled from the geometric response to transverse waves, leading to a symmetric antenna pattern. For $f = 100$ mHz, the asymmetry due to the transfer function becomes apparent.} 
  \label{fig:antenna_pattern}
\end{figure*}

The phase and amplitude response of a detector to a gravitational wave contains information about the relative geometry of the link and the source location. In this section, we will derive the detector response for Doppler-tracking in a one-way photon link to a quasi-monochromatic gravitational-wave \cite{OGderivation,living_reviews_2017, Cornish_2001}.

In the transverse, traceless (TT) gauge, the metric of spacetime is given by 

\begin{equation}
    ds^2 = -c^2dt^2 + dx_idx^i + dx^idx^jh_{ij}(t,\mathbf{x}),
\end{equation}

where ${i,j}$ enumerate the spatial dimensions and $h_{ij}(t,\mathbf{x})$ is the metric perturbation associated with the gravitational wave.

We want to find the coordinate time of arrival for a photon traveling along the direction of the link, whose orientation we can denote by the unit vector $\hat{n}$. Light travels on null geodesics, which are described by $ds^2 = 0$. Expanding the metric perturbation to the first order in $h_{ij}$ gives us, 

\begin{equation}
    c dt = dr + \frac{1}{2}\frac{dx_idx_j}{dr}h_{ij}(t,\mathbf{x}),
\end{equation}

where $dr = \sqrt{dx^2+dy^2+dz^2}$ simply corresponds to the unperturbed photon path. We can further substitute $dx^i = n^idr$, giving us,

\begin{equation}
    c dt = dr \left(1+ \frac{1}{2}n^in^jh_{ij}(t(r),\mathbf{x}(r))\right),
\end{equation}

In the TT gauge, the coordinates of the test masses are fixed at $\mathbf{r}=0$ and $\mathbf{r}=L\hat{n}$, where $L$ is the length of the link. We can integrate this to obtain the coordinate time of arrival for a photon traveling along the link, 

\begin{equation}
    t_{B}-t_{A} = \frac{L}{c} + \frac{1}{2c}\int_{0}^{L} n^in^jh_{ij}(t(r),\mathbf{x}(r)) dr
\end{equation}
Assuming that the gravitational wave is a plane wave with unit wave-vector $\hat{k}$, we can write $h(t,\mathbf{x}) = h(t(r)-\hat{k}.\mathbf{x}(r)/c)$, where $t(r)$ can be written as $t(r) \approx t_B - (L-r)/c$ and $\mathbf{x}(r) = \mathbf{x}(0)+ r\hat{n}$. Using these substitutions, the above integral can be re-written as 

\begin{equation}
    t_{B}-t_{A} = \frac{L}{c} + \frac{1}{2(1-\hat{k}\cdot\hat{n})}\int_{t_B-\frac{L}{c}(1-\hat{k}\cdot\hat{n})}^{t_B} n^in^jh_{ij}(t') dt',
\end{equation}

where we have changed variables in the integral to retarded time 

\begin{equation}
t' = t_B-\left(\frac{L-r}{c}\right)-\frac{\hat{k}}{c}.(\mathbf{x}(0)+r\hat{n}).
\end{equation}

Now, in order to calculate the frequency shift measured at the receiving end (B) of the detector, we just need to take a derivative of the photon arrival time, which is being modulated by the gravitational wave,

\begin{equation}
\begin{aligned}
s(t)
&\equiv \frac{\Delta \nu(t)}{\nu}
= -\frac{\mathrm{d} (t_B-t_A)}{\mathrm{d} t} \\
&= -\frac{1}{2}\,
\frac{n^i n^j}{1 - \hat{k}\!\cdot\!\hat{n}}
\left[
h_{ij}(t)
-
h_{ij}\!\left(
t - \frac{L}{c}\bigl(1 - \hat{k}\!\cdot\!\hat{n}\bigr)
\right)
\right],
\end{aligned}
\label{eq:s(t)_raw}
\end{equation}
which, for a quasi-monochromatic gravitational wave can be reduced to
\begin{equation}
    s(t) =-\frac{1}{2}\frac{n^in^jh_{ij}(t)}{1-\hat{k}\cdot\hat{n}}.\left(1-e^{-i2\pi f \frac{L}{c}(1-\hat{k}\cdot\hat{n}) }\right).
     \label{eq:s_monochrome}
\end{equation}

We can expand $h_{ij}$ into its components $h_{ij} = A_+(t)e^+_{ij}+A_\times(t) e^\times_{ij}$, where $e_{ij}^+$ and $e_{ij}^\times$ are the polarization basis tensors in the detector coordinate frame. This yields 

\begin{equation}
s(t) = F_+A_+(t)+F_\times A_\times(t),
\end{equation} 

where, 

\begin{equation}
    \Fresp = \frac{1}{2}\frac{n^in^je^{+/\times}_{ij}}{1-\hat{k}\cdot\hat{n}}.\left(1-e^{-i2\pi f \frac{L}{c}(1-\hat{k}\cdot\hat{n}) }\right),
    \label{eq:antenna_response}
\end{equation}

is the detector's frequency dependent antenna response. The pre-factor is a purely geometric term obtained by contracting the link vectors with the polarization tensor. The second term corresponds to the detector's transfer function which depends on the relative orientation of the link and gravitational wave propagation vector, as well as the gravitational wave frequency. Here, we have also flipped the sign of the expression with respect to \cref{eq:s_monochrome} since it only contributes to an overall factor of global phase. 

In order to analyze \cref{eq:antenna_response}, we must establish the geometry of the gravitational wave signal relative to the detector frame. As shown in \cref{fig:geometry}, the sky-location of the gravitational wave source can be described by spherical coordinates $\theta$ and $\phi$. Using these coordinates, we can also derive the unit wave-vector $\hat{k}$. In the detector frame, we can define a polarization basis as follows 

\begin{equation}
e_x = \frac{\hat{z} \times \hat{k}}{|\hat{z} \times \hat{k}|}, \qquad e_y = \hat{k} \times e_x,
\end{equation}

The resulting polarization basis tensors are given by

\begin{equation}
 e^+_{ij} = e_x\otimes e_x  - e_y \otimes e_y,\quad e^\times_{ij} = e_x\otimes e_y+e_y \otimes e_x.
\end{equation}

These are rotated from the source's `natural' basis (defined by the inclination angle, $\iota$, between the angular momentum $\hat{L}$ and $\hat{k}$ for a binary) by a polarization angle $\psi$. 

\begin{equation}
    e'^+_{ij} = e^+_{ij} \cos{2\psi}+ e^\times_{ij} \sin{2\psi} 
\end{equation}

\begin{equation}
    e'^\times_{ij} = -e^+_{ij} \sin{2\psi}+ e^\times_{ij} \cos{2\psi} 
\end{equation}

Using these coordinates, we can use \cref{eq:antenna_response} to compute the antenna response for an arbitrarily oriented one-way link. \cref{fig:antenna_pattern} shows the antenna response for a link oriented in the $+x$ direction at two different frequencies, 1 mHz and 100 mHz. At low frequencies, where $2\pi f L/c\ll1$, we can expand \cref{eq:antenna_response} to the leading order 

\begin{equation}
    \Fresp = \frac{1}{2}n^in^je^{+/\times}_{ij}(i 2\pi f L/c),
    \label{eq:antenna_response_lf}
\end{equation}

\begin{figure}
    \centering
    \includegraphics[width=\linewidth]{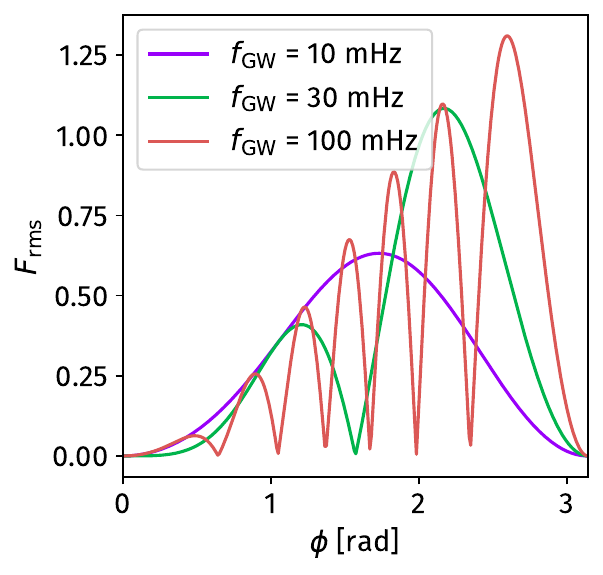}
    \caption{Detector response for $\theta = \pi/2$.  The strong asymmetry between the co-propagating (forward) and counter-propagating (backward) directions is driven by the transfer function. The maximum response occurs when the photon and gravitational wave are nearly aligned ($\hat{k} \cdot \hat{n} \approx 1$), demonstrating coherent phase accumulation. A strict null exists exactly at $\hat{k} = \hat{n}$ due to the transverse nature of the metric perturbation.}
    \label{fig:antenna_in_plane}
\end{figure}

At high frequencies, $f L/c>1$, the transfer function strongly modulates the antenna response, leading to a stark asymmetry between co-propagating and counter-propagating orientations. This directionality is a unique signature of one-way tracking links, as such asymmetries are averaged out in the two-way, round-trip measurements of detectors like LIGO. To see this effect more clearly, we can restrict our signals to the equatorial plane, $\theta = \pi/2$  (see \cref{fig:antenna_in_plane}). 
Notably, when the photon and gravitational wave-vectors are nearly aligned ($\hat{k} \cdot \hat{n} \approx 1$), the response is significantly enhanced despite the transverse nature of the gravitational wave. This occurs because the photon effectively ``surfs" the wave \cite{surfing,surfing2}, with the wavefront traveling at almost the same speed along the link and thus remaining in a region of constant gravitational phase, allowing the stretched or squeezed space to be sampled coherently over the entirety of its journey.

\begin{figure}
    \centering
    \includegraphics[width=\linewidth]{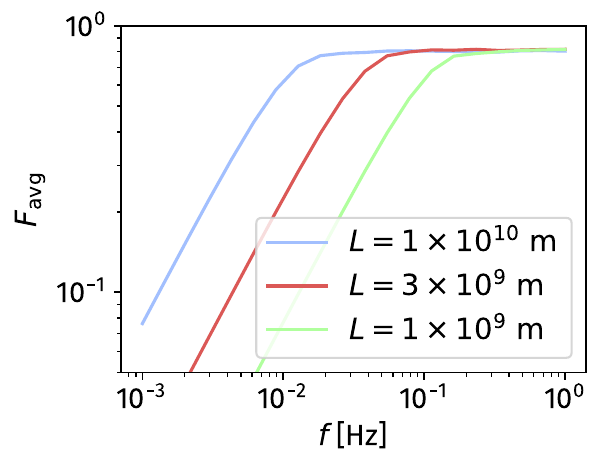}
    \caption{Sky and polarization averaged frequency response of a one-way optical link. The choice of detector baseline length sets a lower limit on frequency band as the detector has a suppressed response for $f<c/L$, while the sensitivity at high frequencies is constant.}
    \label{fig:placeholder}
\end{figure}

\section{Detector Noise}
\label{sec:detector_noise}
\subsection{Quantum Projection Noise}
For a single clock measurement with a finite number, $N$, of atoms and interrogation time $T$, the frequency variance due to quantum projection noise (QPN) is given by 
\begin{equation}
\sigma_f^2  = \frac{1}{NT^2}
\label{eq:QPN}
\end{equation}

In our analysis, we assume an atom number $N = 10^7$. While atom numbers of this magnitude in one-dimensional optical lattices typically introduce density-dependent frequency shifts, the time-varying fluctuations of these shifts, which are the primary concern for AC signals in the gravitational-wave band, can be heavily suppressed through precise stabilization of the lattice intensity and frequency. Furthermore, transitioning to three-dimensional optical lattices can accommodate large atom numbers while virtually eliminating these collisional density shifts by isolating individual atoms.

We also note that while spin squeezing has been successfully realized in optical clocks \cite{SpinSqueezingKaufman,SpinSqueezingYe}, utilizing this technique to surpass the standard quantum limit at state-of-the-art absolute sensitivities remains an ongoing experimental challenge. Therefore, we do not assume spin-squeezed enhancements in this framework.

\subsection{Laser Noise}
\label{sec:laser_noise}

For an optical link receiving finite power $P$, the photon shot-noise on the phase-locked loop results in a residual frequency noise in clock measurements. The power spectrum of such a phase locked loop with control bandwidth $B$ is given by \cite{Kolkowitz16},

\begin{equation}
   S_f(f) =  \frac{(2 \pi f)^2}{(2\pi f)^2+B^2}\left(\Delta_L + \frac{h\nu B^2}{\eta P } \right),
   \label{eq:Sf_laser}
\end{equation}
where $\Delta_L$ is the linewidth of the lasers and $\eta$ is the detection efficiency of the receiving satellite. 

The phase variance due to laser noise in a clock-sequence is given by \cite{Bishof_OSA},  

\begin{equation}
    \sigma_f^2  = \frac{1}{T^2}\int_0^{\infty}df |\tilde{H}(f)|^2S_f(f),
    \label{eq:sigma_f_general}
\end{equation}

where the filter function $\tilde{H}(f)$ is the Fourier transform of the sensitivity (window) function of the sequence. Assuming instantaneous pulses and an optimal bandwidth $B = \sqrt{\eta P \Delta_L/(h\nu)}$, we get the following expression for the frequency variance for a Ramsey sequence of duration $T$, 

\begin{equation}
 \sigma_f^2 = \frac{1}{T^2}\sqrt{\frac{h \nu \Delta_L}{P \eta}},   
 \label{eq:laser_noise_alone}
\end{equation}

An explicit derivation for Ramsey and spin-echo sequences, and the subsequent generalization to dynamical decoupling sequences is described in Appendix \ref{laser_appendix}. For general dynamical decoupling sequences, the laser noise is increased by a factor of $2N_P+1$, where $N_p$ is the number of $\pi$-pulses in the sequence. As one would expect, the noise reduces with decreasing laser linewidth, and increased optical power and detection efficiency. The power received at the end of the satellite link is related to the transmitted power $P_0$ as $P = P_0(\pi R^2 \nu / Lc)^2$, where $R$ is the radius of the emitting and receiving telescopes \cite{Larson2000}. 

The above estimates assume the Rabi pulses are effectively instantaneous. In practice, utilizing Rabi pulses with finite-frequency  $\Omega$ adds a high-frequency roll-off to the sequence's filter function, significantly suppressing the residual laser noise. We numerically evaluate the exact filter functions for finite Rabi pulses in Appendix \ref{laser_appendix}. As shown in \cref{fig:finite_rabi_noise_echo_only}, for $\Omega \ll B$, the noise is greatly suppressed compared to the instantaneous case for spin-echo sequences.

\begin{figure}
    \centering
    \includegraphics[width=\linewidth]{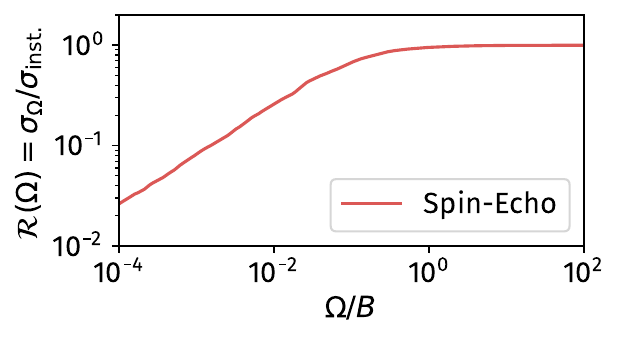}
    \caption{Laser noise suppression from finite Rabi frequency pulses for $\Omega\ll B$. The x-axis has the Rabi frequency $\Omega$ normalized to the laser noise bandwidth $B$, which is set to 70 kHz. The total sequence time $T$ is 100 s.}
    \label{fig:finite_rabi_noise_echo_only}
\end{figure}

Leveraging this finite-pulse noise suppression, we outline three plausible detector scenarios in \cref{tab:laser_noise}, ranging from the most optimistic (Case I) to the least optimistic (Case III). The received optical power in our configurations ($1.2–200$ nW) is optimistic compared to LISA ($\sim$100 pW) \cite{Batkis2025_LISATelescope}. This higher power requires larger telescope apertures, narrowing the beam divergence and tightening absolute pointing requirements. However, because our detection scheme uses a one-way phase lock rather than a round-trip link, the phase-noise budget is decoupled from return-path jitter, potentially relaxing tilt-to-length coupling constraints. Additionally, hardware demands can be traded off by reducing the satellite baseline; this sacrifices low-frequency bandwidth but increases received power as $\propto 1/L^2$, easing the overall link budget.

\begin{table}
    \caption{Laser-noise scenarios for an optical link of length $L = 10^{10}\,\mathrm{m}$, ranging from most (I) to least (III) optimistic. The noise is calculated for a spin-echo sequence, with the finite-pulse noise suppression $\mathcal{R}(\Omega)$ evaluated numerically at each case's parameters (see \cref{fig:finite_rabi_noise_echo_only}). The efficiency $\eta$ is end-to-end, comprising optical throughput, pointing loss, and detection efficiency. The final row gives the frequency below which the detector remains quantum-projection-noise limited, evaluated for $N = 10^7$ atoms and $T_\text{max} = 100$~s.
    }
    \label{tab:laser_noise}
    \begin{ruledtabular}
        \begin{tabular}{lccc}
            \textbf{Parameter}
            & \textbf{Case I}
            & \textbf{Case II}
            & \textbf{Case III} \\
            Emitted power $P_0$
            & $1\,\mathrm{W}$
            & $1\,\mathrm{W}$
            & $100\,\mathrm{mW}$ \\
            Telescope radius $R$
            & $1\,\mathrm{m}$
            & $0.56\,\mathrm{m}$
            & $0.5\,\mathrm{m}$ \\
            Received power $P$
            & $200\,\mathrm{nW}$
            & $20\,\mathrm{nW}$
            & $1.2\,\mathrm{nW}$ \\
            Laser linewidth $\Delta_L$
            & $10\,\mathrm{mHz}$
            & $20\,\mathrm{mHz}$
            & $30\,\mathrm{mHz}$ \\
            End-to-end efficiency $\eta$
            & $0.7$
            & $0.5$
            & $0.3$ \\
            Optimal bandwidth $B$
            & $70\,\mathrm{kHz}$
            & $26\,\mathrm{kHz}$
            & $6.2\,\mathrm{kHz}$ \\
            Rabi frequency $\Omega/2\pi$
            & $10\,\mathrm{Hz}$
            & $10\,\mathrm{Hz}$
            & $5\,\mathrm{Hz}$ \\
            Noise suppression $\mathcal{R}(\Omega)$
            & $0.031$
            & $0.050$
            & $0.073$ \\
            
            \textbf{Phase noise (spin-echo) $\sigma_\phi$}
            & \textbf{$20\,\mu\mathrm{rad}$}
            & \textbf{$75\,\mu\mathrm{rad}$}
            & \textbf{$278\,\mu\mathrm{rad}$} \\
            \textbf{QPN-limited below}
            & \textbf{$1.8\,\mathrm{Hz}$}
            & \textbf{$0.14\,\mathrm{Hz}$}
            & \textbf{$15\,\mathrm{mHz}$}
        \end{tabular}
    \end{ruledtabular}
\end{table}

\subsection{Acceleration Noise}

Data from LISA pathfinder has been used to model acceleration noise as the sum of a Brownian noise background and a $1/f^2$ excess at low frequencies \cite{LISAaccelerationnoise}. We can write the PSD of this noise as,

\begin{equation}
    S_a  = S_b+S_e,
\end{equation}

where $S_b  \approx 3\times10^{-30}$ (m s$^{-2})^2/$Hz, and $S_e \approx 1/f^2\times10^{-36}$ (m s$^{-3})^2/$Hz. The resulting PSD for the Doppler signal is given by 
\begin{equation}
    S_s = S_a/(2\pi f c)^2.
    \label{eq:acc_noise_Ss}
\end{equation}

\begin{figure}
    \centering
    \includegraphics[width=\linewidth]{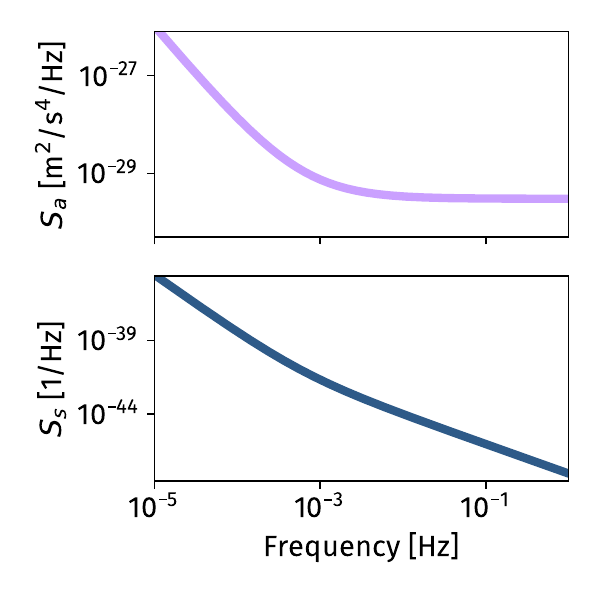}
    \caption{Power spectral density of acceleration noise modeled on LISA Pathfinder data \cite{LISAaccelerationnoise}. The upper subplot shows the PSD in units of acceleration, where we see a Brownian noise floor with an excess $1/f^2$ component at low frequencies. The lower subplot is in units of the measured Doppler signal. }
    \label{fig:acc_noise}
\end{figure}

These PSDs are plotted in \cref{fig:acc_noise}. We note that this model is derived from the differential acceleration of two test masses within a single LISA Pathfinder spacecraft, where common-mode platform motion cancels; our reference masses sit in independent spacecraft, and we therefore treat these values as an optimistic envelope. Evaluating \cref{eq:acc_noise_Ss} at 10~mHz gives an equivalent Doppler noise of $\sqrt{S_s}\approx9\times10^{-23}$~Hz$^{-1/2}$, more than two orders of magnitude below the QPN-limited floor. A tenfold degradation in the acceleration amplitude spectral density, which comfortably covers the loss of common-mode rejection, leaves acceleration noise subdominant. Radiation pressure from the transmitted and solar beams acts on the spacecraft rather than the free-falling reference mass and is compensated by the drag-free loop; such contributions are in any case contained within the measured in-flight residual acceleration of LISA Pathfinder \cite{LISAaccelerationnoise,femtoG_pathfindfer}. The metrology light incident on the mass itself is at the microwatt level, and its fluctuations are negligible against the Brownian floor for any reasonable degree of power stabilization.

\subsection{Time-Varying Systematic Shifts}

Time-varying systematic shifts of the atomic clock transition frequencies set a noise floor below which gravitational wave signals cannot be detected. Although the accuracies of state-of-the-art optical lattice clocks are limited by  uncertainty regarding absolute frequency shifts, primarily driven by blackbody radiation, magnetic fields, and lattice light shifts, static DC offsets do not affect the measurement of transient gravitational waves. The sensitivity of the detector is instead constrained by how these environmental parameters fluctuate over time within the specific signal frequency band, filtered by the detection pulse sequence. 

For a rough estimate of the required stability of these environmental parameters, we can evaluate the first-order, in-band fluctuations necessary to keep systematic fractional frequency variations below the $10^{-22}$ level at the target signal frequencies.

The blackbody radiation frequency shift for~\Sra~is given by \cite{BBRShift,BBRShiftOG},
\begin{equation}
 \Delta\nu/\nu  \approx -5\times10^{-15}\left(\frac{T}{300\text{K}}\right)^4,  
\end{equation}
where $T$ is the blackbody temperature of the clock environment. This yields a sensitivity of, 
\begin{equation}
d(\Delta\nu/\nu)/dT  \approx -6.6\times10^{-17}\left(\frac{T}{300\text{K}}\right)^3   
\end{equation}
Recently, radiation shielding in optical lattice clocks has been shown to achieve nearly perfect blackbody environments at cryogenic temperatures \cite{CryoClock}. At 77 K, differential temperature fluctuations at signal frequencies must be suppressed to the order of $0.1$ mK, in order to keep the resulting fractional frequency fluctuations below $10^{-22}$. This requirement could be further relaxed by operating at even lower temperatures. 

For magnetic fields, the Zeeman shift poses a similar constraint. In $^{87}\text{Sr}$, the $|g,5/2\rangle\leftrightarrow|e,3/2\rangle$ clock transition exhibits a linear magnetic field sensitivity of $2.24\times10^{5}\,\text{Hz/T}$ \cite{Zheng22}. To ensure that magnetic-field-related fractional frequency fluctuations remain below $10^{-22}$ in the signal band, differential magnetic field fluctuations must be controlled to roughly $0.1\,\text{pT}$. However, first-order magnetic field fluctuations can be nearly completely eliminated by interrogating interleaved atomic ensembles on both the $|g,5/2\rangle\leftrightarrow|e,3/2\rangle$ and $|g,-5/2\rangle\leftrightarrow|e,-3/2\rangle$ transitions. Averaging the two frequencies cancels out the linear Zeeman shift because the shift has opposite signs for the symmetric $m_F$ projections \cite{LudlowReview,Zheng23,Ohmae:20}. 

The residual systematic noise is then dominated by the much weaker quadratic Zeeman shift, given by \cite{SecondOrderZS}:$$\Delta\nu \approx -0.23\left(\frac{B}{1\,\text{G}}\right)^2\,\text{Hz}.$$In the presence of a bias field of $1\,\text{G}$, reducing the quadratic-Zeeman-induced fractional frequency fluctuations below $10^{-22}$ requires the in-band magnetic field fluctuations to be suppressed to the level of $10\,\text{pT}$.

In optical lattice clocks, we must also account for the light shift from the  lattice light that is used to strongly confine the atoms. To mitigate this effect, lattice clocks operate at a `magic wavelength,' where the shift in ground and excited states are identical to leading order. However, frequency noise in the lattice laser can lead to an associated time-varying frequency shift. The light shift is given by \cite{MagicIntensity,Zheng23}, 

\begin{equation}
 \Delta\nu_{\text{LS}}(\delta_L,u,\delta u )  \approx  \left[\left(\frac{\partial\alphaE}{\partial\nu}\delta_L- \alphaqm\right)\frac{\sqrt{u}}{2} - \frac{\partial\alphaE}{\partial\nu}\delta_Lu\right],
 \label{eq:lattice_shift}
\end{equation}

where $\delta_L$ is the detuning of the lattice from the magic wavelength and $u$ is the trap depth normalized to the lattice photon recoil energy $E_R$. $\alphaE$ and $\alphaqm$ are the differential $E1$ and $E2$-$M1$ polarizability of the clock transitions (the values of these parameters can be found in Ref. \cite{MagicIntensity}).  

For~\Sra~lattice clocks operating at a depth of around $10 E_r$ in an 813 nm lattice, the fluctuations in the lattice frequency must be suppressed to the order of 300 Hz in the signal band in order to meet the $10^{-22}$ requirement. From \cref{eq:lattice_shift}, we see that fluctuations in the trap depth can also lead to time varying shifts due to the higher order $E2$-$M1$ polarizability. At the magic wavelength, the fluctuations in the trap depth/lattice intensity must be controlled to a part in $10^4$ in the signal band. At lattice detunings $>10$ MHz from magic, the contribution from the $E1$ term becomes significant, leading to stricter intensity control requirements. 

Finally, the clock and lattice lasers must be referenced to the drag-free proof mass,
and any relative motion along this chain,  arising from the local metrology, servo
residuals, fiber paths, or the mounting of the lattice retro-reflector, is
indistinguishable from a gravitational-wave-induced Doppler shift. A path-length
fluctuation $\delta x$ at frequency $f$ produces a fractional frequency shift
$\delta\nu/\nu \approx 2\pi f\,\delta x/c$, so keeping the resulting fractional
frequency fluctuations below $10^{-22}$ at 10 mHz requires in-band path-length
fluctuations across the full referencing chain to be held below $\approx0.5$ pm. This
is comparable to the picometer-level optical metrology requirements of LISA
\cite{LISA_Pathfinder}, and represents one of the principal technological requirements
for a detector of this kind.

\subsection{Strain Sensitivity}

\begin{figure}
    \centering
    \includegraphics[width=\linewidth]{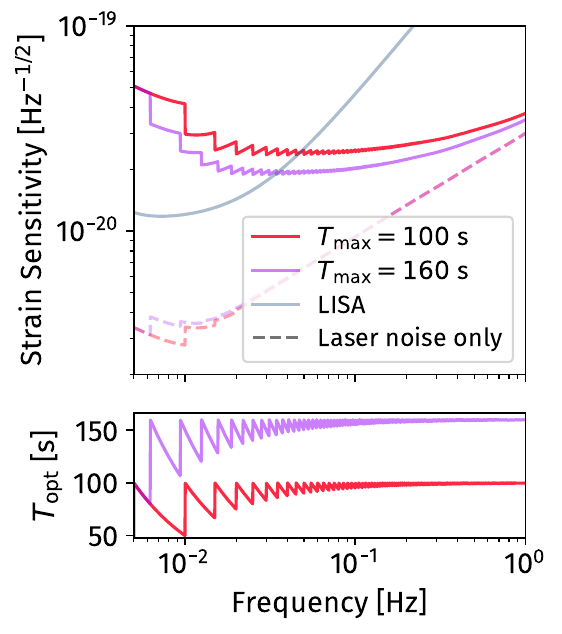}
    \caption{Strain sensitivity for a clock-based gravitational wave detector for a detector with $L = 10^{10}$ m. The strain sensitivity of LISA \cite{Robson_2019} is plotted alongside for comparison. $T_\text{max}$ is the maximum allowed interrogation time for a clock pulse sequence, set by atomic coherence time. $T_{\text{opt}}$ is the optimal interrogation time for a given signal frequency. Dashed traces show the residual laser noise contribution alone, using Case I of \cref{tab:laser_noise}. Note that this curve is a locus of noise performance over tuning \cite{LSD_OG,LSD,GW_res}. A signal whose frequency is not known in advance is instead searched for as described in \cref{sec:search_mode}.}
    \label{fig:PSDs}
\end{figure}
Given the negligible effect of acceleration noise, while  considering the total noise for an optimized spin-echo measurement at frequency $f$, we only include contributions from QPN and residual laser noise:

\begin{equation}
    \sigma^2(f,\tau) = \frac{1}{(2\pi\nu)^2T(f)\tau}\left(\frac{1}{N}+|\mathcal{R}(\Omega)|^2(2N_P+1)\sqrt{\frac{h\nu\Delta_L}{P\eta}}\right)
    \label{eq:total_noise}
\end{equation}

Here, the interrogation time $T$ is chosen to match the maximum possible number of gravitational wave half cycles that can be fit within the maximum possible interrogation time $T_{\text{max}}$, which is set by the atomic coherence. $\tau$ is the averaging time

\begin{equation}
    T_\text{opt} = \lfloor 2fT_{\text{max}}\rfloor /2f, 
    \label{eq:T_opt}
\end{equation}

 where $f$ is the frequency of the measured signal. The number of $\pi$-pulses, $N_p$ required for an appropriate dynamical decoupling sequence is also related as, 

 \begin{equation}
     N_P = \lfloor 2fT_{\text{max}}\rfloor-1.
     \label{eq:N_p}
 \end{equation}

Using $N = 10^7$ atoms and the optimistic laser noise case (I) from \cref{tab:laser_noise}, we plot an effective maximum strain sensitivity in \cref{fig:PSDs} for a detector with a $10^{10}$ m baseline, along with the optimal interrogation times as a function of frequency, for two different choices of $T_{\text{max}}$. Note that this sensitivity can only be achieved when the detector is exactly tuned to the signal frequency as is typically the case with resonant detectors \cite{LSD_OG,LSD, GW_res}. The plotted strain noise is the locus of noise performance over the detection band.

As expected from the detector response, the sensitivity is reduced at frequencies where $f<c/L$. At higher frequencies, there is an increase in noise associated with the high number of $\pi$-pulses $N_p$ needed for the appropriate dynamical decoupling sequence. As with Ramsey and spin-echo sequences, finite Rabi frequency noise suppression can be achieved with dynamical-decoupling sequences too. We leave a more detailed study of the impacts of laser phase noise in dynamical decoupling sequences with many pulses for future work. Throughout the remainder of the analysis in this paper, we will assume detector noise corresponding to $T_{\text{max}}=100$ s.

\section{Phase and Amplitude Estimation Protocol}

\label{sec:protocols}
We assume that the Doppler signal as measured by the detector is quasi-monochromatic with frequency $f$ with a fixed amplitude $A$ and phase $\Phi$,
\begin{equation}
    s(t) = A\sin(2\pi f t+\Phi)
\end{equation}

The phase and amplitude of the measured signal are related to the gravitational wave strain via the detector's response function $\Fresp$, defined in \cref{eq:antenna_response}. In this measurement protocol, we assume that the frequency of the signal is known beforehand, using information from LISA \cite{CMI}. We use a spin-echo sequence of length $T = 1/f$ starting at time $t_0$. A similar analysis can be done for both Ramsey and dynamical-decoupling sequences. The total measured phase is given by 

\begin{equation}
    \overline{s} = \frac{1}{T} \int_{t_0}^{t_0+T}s(t) y(t)dt,
\end{equation}

where,

\begin{equation}
    y(t) =
\begin{cases}
1 & t_0<t < t_0+T/2, \\
-1  &  t_0+T/2<t<t_0+T,
\end{cases}
\label{eq:spin-echo-window}
\end{equation}
is the window function for the spin-echo echo sequence. This yields

\begin{equation}
    \overline{s} = \frac{2A}{\pi}\cos(2\pi f t_0 +\Phi).
    \label{eq:phase_est}
\end{equation}

\begin{figure}
    \centering
    \includegraphics[width=\linewidth]{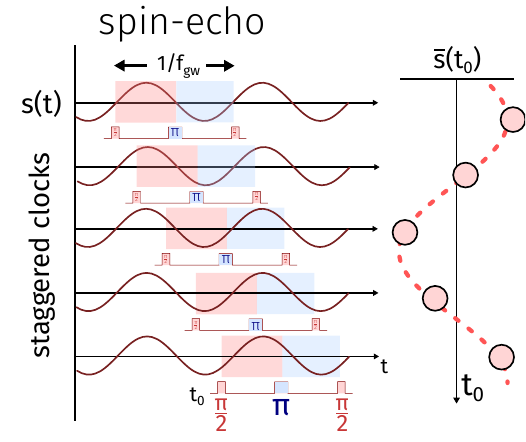}
    \caption{Phase/Amplitude Estimation Protocol. Spin-echo sequences with multiple atomic ensembles (clocks) in a single detector are used measure gravitational wave signals with a known frequency. The resulting measurements are then plotted as a function of measurement start-time in order to extract the phase and amplitude of the detected GW signal using \cref{eq:phase_est}. Note that this protocol could be extended to general dynamical decoupling sequences, with the use of an appropriate sign between consecutive measurements in order to average the signal over long times.}
    \label{fig:phase_est}
\end{figure}

By simultaneously measuring this signal with multiple staggered clocks/atomic ensembles (see \cref{fig:phase_est}), beginning at distinct, evenly spaced, $t_0$, and fitting \cref{eq:phase_est} to the measured results, we can estimate the phase and amplitude of the gravitational-wave signal. While only two staggered clocks measuring orthogonal signal quadratures are needed to entirely estimate phase and amplitude, the addition of more staggered clocks improves the signal-to-noise ratio without having to load more atoms in a single ensemble.

In order to reduce the effect of noise, these measurements are averaged over multiple cycles, provided the signal remains largely constant over the averaging time. For a signal with constant frequency, a clock with zero dead-time between measurements, ensures this. Note that for Ramsey measurements, or dynamical decoupling sequences corresponding to an odd number of gravitational wave half-cycles, we must alternate the sign of the signal between consecutive measurements.

We use simulated data and noise models in order to evaluate the performance of this
measurement protocol as a function of measurement parameters. The simulated signal in
both cases corresponds to a binary black-hole inspiral, whose source parameters are
listed in \cref{tab:sim_source}. The starting frequency at which the signal is measured
is $10$ mHz, corresponding to a spin-echo sequence time of 100 s. This simulation takes
the signal chirp into account, using a slightly modified measurement protocol which is
discussed in the next sub-section. For simplicity, we assume that the geometrical antenna
response remains constant over the measurement. Simulated noise is added to each
measurement and maximum-likelihood estimation is used to extract the measured signal.
Since the contrast must be fit jointly with $\phi_D$ (\cref{sec:diff_comp}), the
measurements are grouped into blocks of at least four consecutive shots, over which
$\phi_D$ is treated as constant. This is not restrictive for the slowly chirping signals
considered here: across four sequences at $f = 10$~mHz, the interrogation time, and hence
$\phi_D$ at fixed signal amplitude, changes by only $\sim10^{-4}$, placing the resulting
variation in $\phi_D$ well below the per-shot quantum projection noise.

\begin{table}
    \caption{Source parameters for the simulated binary black hole inspiral signal.}
    \label{tab:sim_source}
    \begin{ruledtabular}
        \begin{tabular}{lc}
            Chirp mass $\Mc$ & $1000\,M_\odot$ \\
            Signal frequency $f$ & 10 mHz\\
            Redshift ($z$) & 0.01\\
            Luminosity Distance $D_L$ & 43 Mpc\\
            Strain amplitude $h$ & $1.3\times10^{-20}$ \\
            Altitude $\theta$ & $45.8^\circ$ \\
            Azimuth $\phi$ & $11.45^\circ$ \\
            Inclination angle $\iota$ & $60^\circ$ \\
            Polarization angle $\psi$ & $60^\circ$
        \end{tabular}
    \end{ruledtabular}
\end{table}

\begin{figure}
    \centering
    \includegraphics[width = \linewidth]{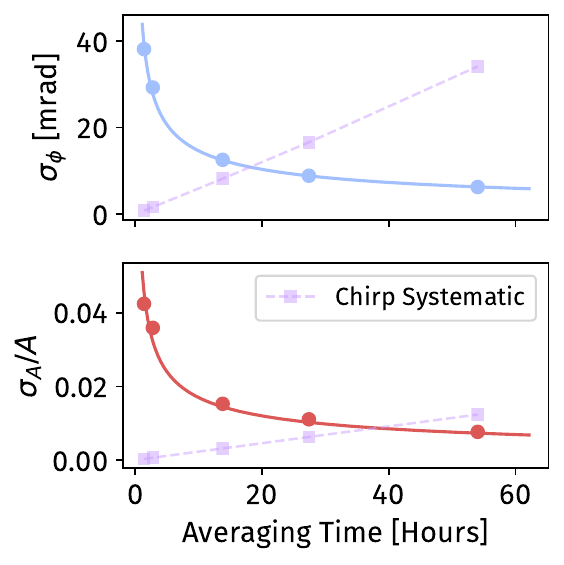}
    \caption{Phase and amplitude estimation uncertainty as a function of averaging time for a binary-black hole inspiral signal with source parameters detailed in \cref{tab:sim_source}. The detector uses 5 staggered clocks, with $N = 10^7$ atoms and laser noise corresponding to Case I in \cref{tab:laser_noise}. The purple dashed traces show the systematic bias from the signal chirp, which grows linearly with
averaging time.}
    \label{fig:phase_amp_averaging_time}
\end{figure}

\begin{figure}
    \centering
    \includegraphics[width = \linewidth]{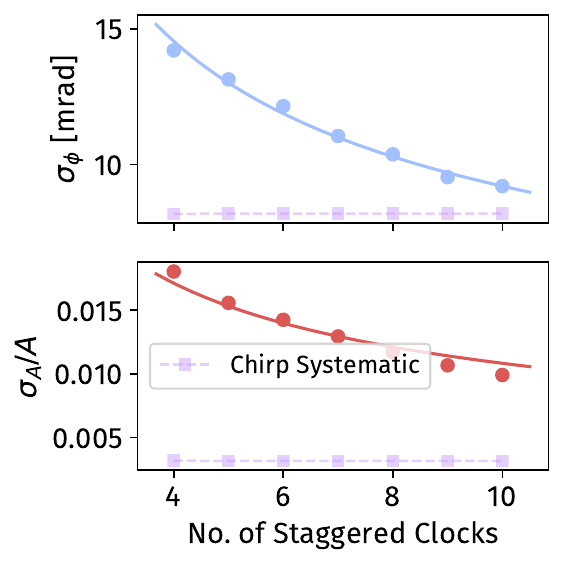}
    \caption{Phase and amplitude estimation uncertainty as a function of the number of staggered clocks for a binary-black hole inspiral signal with source parameters detailed in \cref{tab:sim_source}, for an averaging time of 13.8 hours. The detector has $N = 10^7$ atoms and laser noise corresponding to Case I in \cref{tab:laser_noise}. The purple dashed
traces show the systematic bias from the signal chirp, which is set by the averaging time and is therefore independent of the number of clocks.} 
    \label{fig:phase_amp_n_ensembles}
\end{figure}

\cref{fig:phase_amp_averaging_time} shows the phase and amplitude uncertainties as a function of averaging time for a measurement with 5 staggered clocks. As expected, we see these uncertainties go down with measurement time, following the familiar $1/\sqrt{T}$ trend. In \cref{fig:phase_amp_n_ensembles}, we plot these uncertainties as a function of the number of staggered clocks, for an averaging time of around 13.8 hours, again seeing a trend that roughly corresponds to $1/\sqrt{N}$.

We also note the systematic bias resulting from the effect of the signal chirp on the transfer function in \cref{eq:antenna_response}. Averaging over a time $\tau$ biases the recovered phase and amplitude; the bias grows
linearly with $\tau$, unlike the $1/\sqrt{\tau}$ statistical error, and is not reduced by
adding clocks. For the source in \cref{tab:sim_source} the two are comparable at
$\sim19$ hours. This bias is also present in the network measurements of \cref{sec:pe},
where we discuss its implications for parameter estimation.

\newpage

\subsection{Chirping Signals}

\begin{figure}
    \centering
    \includegraphics[width = \linewidth]{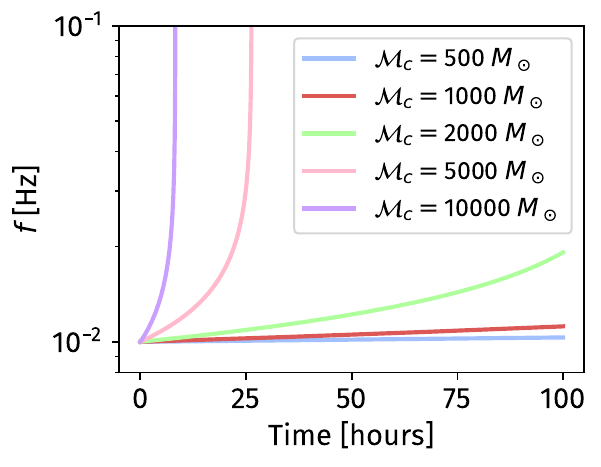}
    \caption{Frequency evolution of gravitational waves from compact binary inspirals for various chirp masses, with $f = 10$ mHz at $t=0$.}
    \label{fig:inspiral_chirps}
\end{figure}

The above protocol is ideal for signals with a stationary, constant frequency. However, typical gravitational wave signals from compact binary mergers chirp at a rate given by

\begin{equation}
    \dot{f} = \frac{96}{5}\pi^{\frac{8}{3}}\left(\frac{G \Mc}{c^3}\right)^{\frac{5}{3}}f^{\frac{11}{3}}.
    \label{eq:chirp}
\end{equation}
In \cref{fig:inspiral_chirps}, we plot this frequency evolution for a variety of binary
chirp masses. We see that for heavier binaries, the signal chirps extremely rapidly,
allowing for almost no averaging time within the measurement band. In this paper, we
restrict our analysis to slowly chirping signals so that they can be effectively averaged
over the measurement time, without introducing severe biases from the frequency dependence
of the detector response. This also sets our choice of benchmark: at $f = 10$ mHz a $\Mc = 1000 M_\odot$ binary is 
$\sim16$ days from merger, while the same system at 100 mHz merges in $\sim0.8$ hours. 
The benchmark is chosen as a regime where the protocol can be demonstrated end to end, 
not one where the detector is most competitive.

\begin{figure}
    \centering
    \includegraphics[width=\linewidth]{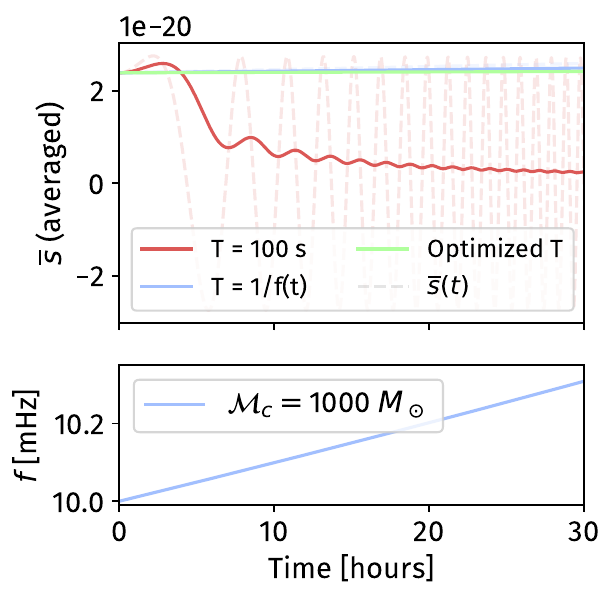}
    \caption{Effect of signal chirp on an averaged frequency measurement}
    \label{fig:chirp_average}
\end{figure}

To ensure consistent signal averaging, the interrogation time
must be dynamically updated for every measurement. In our analysis, we numerically solve
for the interrogation time $T$ that spans one cycle of the signal,
\begin{equation}
    \int_t^{t+T}f(t')dt' = 1.
\end{equation}
For compact binaries, the relationship between $f$ and $t$ is well modeled, allowing us to
easily make this computation provided we have prior knowledge of the chirp mass.
\cref{fig:chirp_average} shows how a measured signal averages over time for different
choices of interrogation times. We also see that simply tracking the instantaneous
frequency, $T(t) = 1/f(t)$, performs as well as the numerically solved interrogation time
over the averaging times considered here.

\subsection{Signal Priors}
\label{sec:signal_priors}
The protocol described above requires the signal frequency, and its evolution, to be supplied in advance: the frequency sets the interrogation time, and the chirp rate sets how that interrogation time must be updated as the signal evolves. Two independent errors in the supplied model matter. An error $\delta t_c$ in the timing of the signal offsets the frequency at our observing epoch by $\delta f = \dot{f}\,\delta t_c$, while an error in the chirp mass leaves the frequency correct but the chirp rate wrong,

\begin{equation}
	\frac{\delta\dot{f}}{\dot{f}} = \frac{5}{3}\frac{\delta\Mc}{\Mc}.
	\label{eq:dfdot}
\end{equation}

Both accumulate phase against the true signal, so that after a time $t$ the model is in error by $\varepsilon(t) = 2\pi\left[\delta f\,t + \tfrac{1}{2}\delta\dot{f}\,t^{2}\right]$. Since the phase is estimated from measurements distributed across the averaging window, the relevant quantity is the mean of this error over the window rather than its value at the end,

\begin{equation}
	\delta\Phi(\tau) = \frac{1}{\tau}\int_0^{\tau}\varepsilon(t)\,dt
	= \pi\dot{f}\,\tau\left[\delta t_c + \frac{5}{9}\frac{\delta\Mc}{\Mc}\,\tau\right].
	\label{eq:phase_error_chirp}
\end{equation}

For the source of \cref{tab:sim_source}, taking $\delta\Mc/\Mc = 10^{-5}$ and $\delta t_c = 10$~s~\cite{CMI} at an averaging time of $\tau = 14$ hours yields $\delta\Phi \approx 4$~mrad, roughly a quarter of the statistical phase uncertainty of $\sigma_\Phi\approx18$~mrad at the same averaging time, so the externally supplied signal model does not limit the measurement.

\subsection{Search Mode}
\label{sec:search_mode}

\cref{fig:chirp_average} suggests a way to operate without a prior on the signal morphology. Even with a fixed interrogation time, the averaged signal builds up for a while before the chirp dephases it. This is enough to support a search mode. With the interrogation time parked at a fixed $T$, the sequence acts as a narrowband filter centered near $f \sim 1/T$, and the detector can be used to search for signals within that band without any prior on the source.

What such a search gives up is coherent averaging time. Without a prior on the chirp mass the interrogation time cannot track the frequency evolution, and the signal dephases from a fixed-frequency template once the accumulated phase error becomes appreciable.
Expanding the signal phase about a fixed frequency, $\Phi(t) = 2\pi\int_0^t f\,dt' \approx 2\pi(f_0 t + \tfrac{1}{2}\dot{f}t^2)$, the error accumulated against a constant-frequency template grows as $\pi\dot{f}t^2$, so that coherence is lost after

\begin{equation}
	\tau_\text{coh} \simeq \frac{1}{\sqrt{\dot{f}}},
	\label{eq:tau_coh}
\end{equation}

\cref{tab:search_snr} lists $\tau_\text{coh}$ and the corresponding loss in signal-to-noise for a representative set of chirp masses at two signal frequencies.

\begin{table}
	\caption{Coherent averaging time available in search mode, and the resulting signal-to-noise relative to a targeted measurement of the same source averaged over 14 hours.}
	\label{tab:search_snr}
	\begin{ruledtabular}
		\begin{tabular}{lcccc}
			& \multicolumn{2}{c}{\textbf{$f = 10$ mHz}}
			& \multicolumn{2}{c}{\textbf{$f = 20$ mHz}} \\
			\cline{2-3}\cline{4-5}
			\textbf{$\Mc\ [M_\odot]$}
			& $\tau_\text{coh}$ & SNR/SNR$_\text{targ.}$
			& $\tau_\text{coh}$ & SNR/SNR$_\text{targ.}$ \\
			\colrule
			$1000$ & $5.0\,\mathrm{h}$ & $0.60$ & $1.4\,\mathrm{h}$ & $0.32$ \\
			$1500$ & $3.5\,\mathrm{h}$ & $0.50$ & $1.0\,\mathrm{h}$ & $0.27$ \\
			$2000$ & $2.8\,\mathrm{h}$ & $0.45$ & $0.8\,\mathrm{h}$ & $0.24$ \\
		\end{tabular}
	\end{ruledtabular}
\end{table}
We stress that this is a deliberately rudimentary search strategy, intended only to establish that the instrument retains sensitivity without a prior. The question of the optimal strategy, together with the associated detection thresholds and false-alarm rates, is a substantial problem in search design that we leave to future work.

\section{Astrophysical Parameter Estimation}
\label{sec:pe}
\begin{figure}
    \centering
    \includegraphics[width=\linewidth]{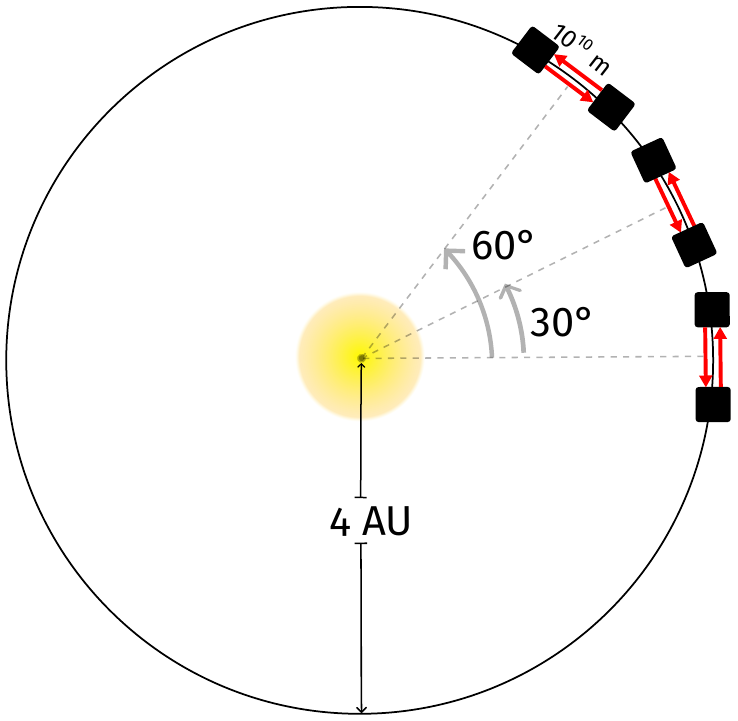}
    \caption{Example clock-detector network, with 6 one-way optical links across 3 satellite pairs. Each link has length $10^{10}$ m. The constellation is in a heliocentric orbit at 4 AU, a radius at which the geometric response of the network is effectively constant over measurement averaging times considered in this work.}
    \label{fig:detector_network}
\end{figure}

\begin{figure*}
    \centering
    \includegraphics[width=\linewidth]{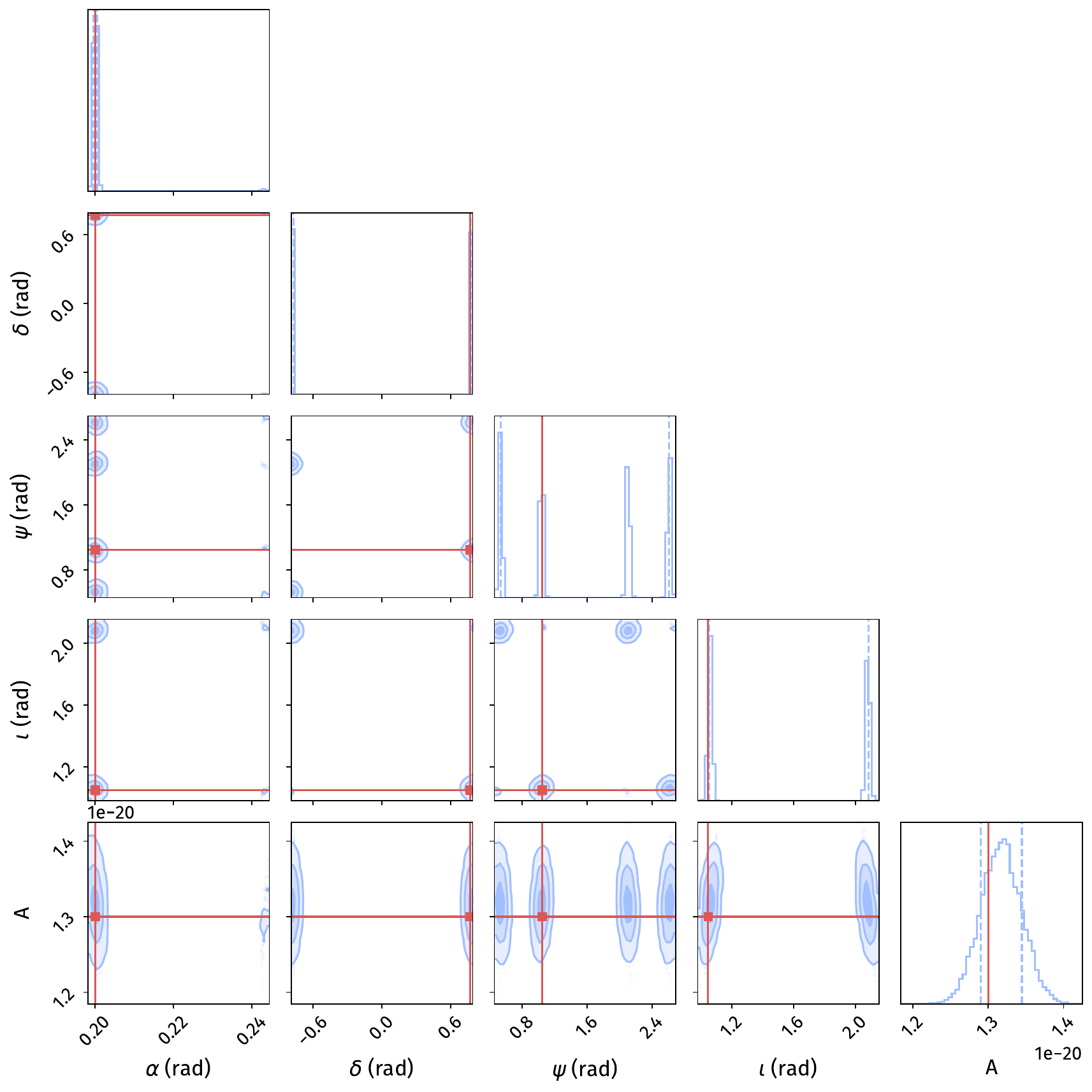}
    \caption{Posterior distributions for simulated binary inspiral signal at $f = 10$ mHz, as measured by a clock network, with data averaged over 27 hours. The source parameters, given in \cref{tab:sim_source}, are marked on the plot in red and clearly lie within the estimated posterior distributions.}
    \label{fig:corner}
\end{figure*}

In this section, we conduct preliminary data analysis on simulated measurements from a clock network, as an example to demonstrate the utility of our measurement framework. In particular, we consider a detector network consisting of 3 pairs of satellites in a heliocentric orbit (\cref{fig:detector_network}). Each baseline contains two one-way links of length $10^{10}$ m, giving us a total of 6 detectors. For simplicity of analysis, we assume that the geometric response of the network does not change significantly over the measurement averaging time. To that effect, the orbit radius is chosen to be 4 AU. In future work, we plan to extend this analysis to a general case of detectors with time-varying antenna responses. Each detector uses 5 staggered clock comparisons to implement the phase and amplitude estimation protocols described in the previous section. The noise model considered is optimistic, with QPN corresponding to $N=10^7$, and residual laser noise given by Case I from \cref{tab:laser_noise}. The relative phase and amplitude response of each detector is used to extract information about the gravitational wave source. 

As a representative science case, we will investigate the sky localization of intermediate-mass black hole (IMBH) systems. While previous work has established foundational limits on gravitational-wave localization using atomic sensors via Fisher information analyses \cite{AILocalization}, this work extends those efforts by performing full parameter estimation on simulated data. By modeling explicit measurement protocols, we are able to extract complete Bayesian posterior distributions for the sky location, allowing us to capture degeneracies and uncertainties beyond linear approximations.

We consider a list of source parameters $\hat\theta = \{\alpha, \delta, \iota, \psi, A\}$ given a set of detector measurements $\mathbf{d}$. Here, following astrophysical convention, the right ascension ($\alpha$), and declination ($\delta$), of the source are related to the spherical coordinates $\theta$ and $\phi$ as $\alpha=\phi$ and $\delta=\pi/2-\theta$. As discussed in \cref{sec:detector_response}, the binary inclination angle $\iota$ and signal polarization $\psi$ are related to the angular momentum vector of the source (see \cref{fig:geometry}).

In this work, we restrict our analysis to a simple model of gravitational waves emitted by compact binary mergers, where the signal contains a single quasi-monochromatic component. For such a signal at frequency $f$, each detector $d$ measures a complex strain response,

\begin{equation}
    S_d = [F_+(\alpha,\delta,\Psi) A_+ + i F_\times(\alpha,\delta,\Psi) A_\times
]\exp(i 2\pi f \tau_D).
\label{eq:complex_response}
\end{equation}

Here the antenna response is given by \cref{eq:antenna_response}. Note that, for a binary merger, the two polarization components are $\pi/2$ out of phase with each other. The response also encodes a relative phase delay due to the finite GW travel time $\tau_D$ between the detectors.  

In the ``source frame", the relative magnitudes of the $+$ and $\times$ polarized components of the gravitational wave are given by 

\begin{equation}
A_+ = A\left(\frac{1+\cos^2({\iota})}{2}\right),\quad \quad A_\times  = A \cos(\iota),
\label{eq:amplitudes}
\end{equation}
where $A$ is the gravitational wave amplitude, which, at frequency $f$, is given by, 

\begin{equation}
A = \frac{4}{c^4} (G\Mc(1+z))^{5/3} \frac{(\pi f)^{2/3}}{D_L(z)},
\label{eq:strain_amp}
\end{equation}

for a binary with chirp mass $\Mc$ at redshift $z$. The luminosity distance, $D_L$, is related to the redshift as, 
\begin{equation}
    D_L(z) = (1+z)c\int_0^z\frac{dz'}{H_0\sqrt{(\Omega_m(1+z')^3+\Omega_\Lambda)}}.
    \label{eq:luminosity_dist}
\end{equation}

Using a Bayesian framework for our analysis, we can write the posterior probability distribution of parameters, given a data set $\mathbf{d}$, as, 

\begin{equation}
    p(\hat\theta| \mathbf{d}) = \frac{\mathcal{L}(\mathbf{d} | \theta)\, \pi(\theta)}{\mathcal{Z}}.
\end{equation}

Here, $\mathcal{L}(\mathbf{d} |\hat\theta)$ is the likelihood of observing the data $\mathbf{d}$ given a particular parameter set $\hat\theta$, $\pi(\theta)$ is the prior probability which encodes our knowledge of the parameters before observing the data, and $\mathcal{Z} = \int \mathcal{L}(\mathbf{d}|\hat\theta)\,\pi(\hat\theta)\,d\hat\theta$ is the Bayesian evidence, which serves as the normalization constant and enables model comparison. The construction of the likelihood function and the choice of Bayesian priors are discussed in Appendix \ref{app:bayesian}.

In \cref{fig:corner}, we plot the posterior probability distribution of source parameters for a simulated source with parameters described in \cref{tab:sim_source}, for a total averaging time of 27 hours. We see that the true signal parameters, marked in red, are within the estimated posterior regions. The mirror-image degeneracy in declination $\delta$ is due to the fact that all the detectors lie in the same plane. 

\begin{figure}
    \centering
    \includegraphics[width=\linewidth]{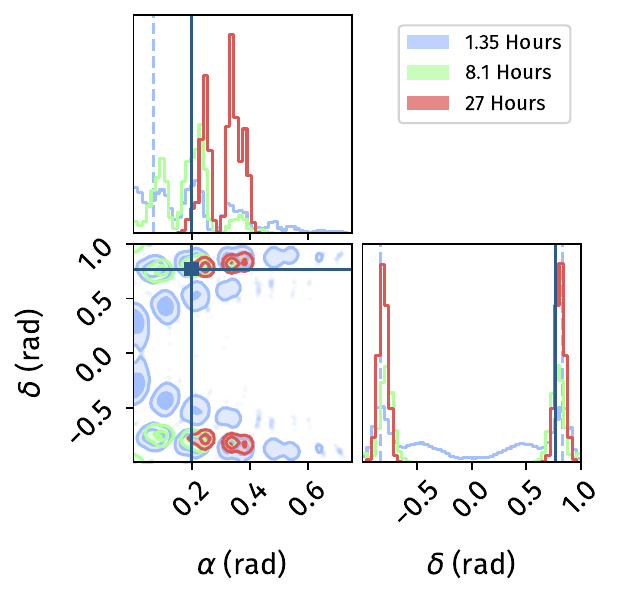}
    \caption{Evolution of sky localization for an binary inspiral at $f = 10$ mHz with $\Mc = 1500 M_\odot$ at $z=0.1$ as the averaging time is increased.}
    \label{fig:corner_evol}
\end{figure}

\begin{figure}
    \centering
    \includegraphics[width=\linewidth]{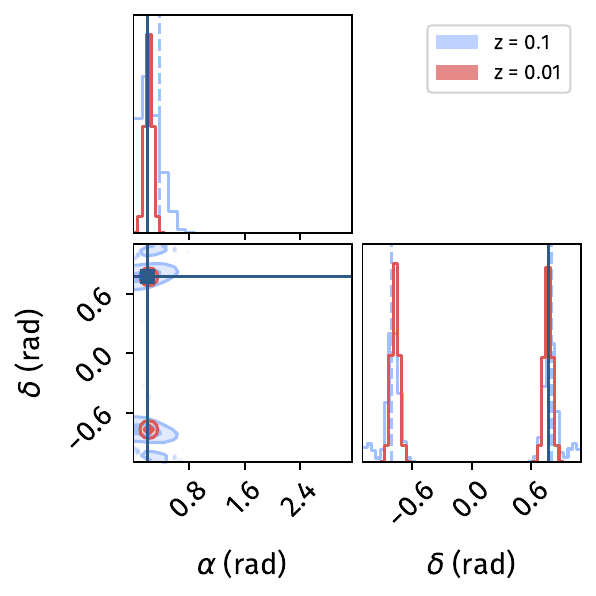}
    \caption{Sky localization posteriors for a binary inspiral at $f  = 10$ mHz with $\Mc = 1000 M_\odot$ at $z=0.1$ and $z=0.01$, with signals averaged over 27 hours.}
    \label{fig:corner_z}
\end{figure}

In \cref{fig:corner_evol}, we explicitly plot the evolution of posterior distributions for sky coordinates, as they evolve with signal averaging time for a $\Mc = 1500M_\odot$ system at $z=0.1$. We see that by increasing the signal-to-noise by averaging for longer, we are able to resolve some of the degeneracies between various disconnected sky-locations. In \cref{fig:corner_z}, we compare the sky location posteriors for the same source at different redshifts. As expected, a larger redshift (smaller amplitude) results in larger uncertainties. In future work, we plan to use information from the frequency evolution of the signal and the orbital motion of the network constellation to further improve on the localization. 

We evaluate source strains from $10^{-21}$ to $10^{-20}$ ($z \approx 0.1$ to $0.01$)
purely as a representative benchmark. In practice, the exact range of resolvable signals
depends on the total measurement time. Although a typical source will stay within the
detector band for an extended period, the frequency evolution of the signal leads to a
time-varying detector response, which requires more sophisticated analysis techniques
such as sequential Bayesian updating using smaller time-binned signal averages. The
resulting bias in the recovered phase and amplitude, quantified in \cref{sec:protocols},
is present in the simulated data analyzed here, and likely contributes to the offsets
visible between the injected parameters and the peaks of the posterior distributions in
\cref{fig:corner,fig:corner_evol,fig:corner_z}. As with the time-varying geometric response, an in-depth analysis of
these dynamics, optimal network configurations, and dynamic parameter estimation
techniques is deferred to dedicated future work.

\section{Conclusions and Outlook}
In this work, we have shown that a space-based clock detector network can open the door to the yet-unexplored $0.01–1$ Hz band in gravitational-wave astronomy through its unique and tunable design. Although our quantitative analysis focused on optical lattice clocks as the primary technological platform, the measurement schemes described are broadly applicable to general optical clock technologies. Specifically, advances in the emerging field of nuclear clocks could prove to be an extremely promising platform for such a network \cite{nuclear_clock}. 

Furthermore, we demonstrate that it is possible to extract crucial astrophysical information about compact binary sources using the aforementioned measurement protocols. As an example case, we conduct preliminary parameter estimation on simulated data to demonstrate sky localization of intermediate-mass black hole (IMBH) mergers. These protocols are targeted, in that they require the signal frequency, and for chirping sources its evolution, to be supplied externally, as would be the case for a source already detected by LISA. In \cref{sec:search_mode} we show that the same instrument can instead be operated as an independent narrowband search at a modest cost in sensitivity, although a full treatment of search design lies beyond the scope of this work. In future work, we plan to extend this analysis to more general scenarios by incorporating the orbital motion of the detector constellation and the time-evolving, frequency-dependent response of these detectors to chirping signals, eventually extending our study to novel science cases in the decihertz regime.

Finally, advancing the technological requirements for these detectors is vital to the broader field of optical clocks, and space-based clock networks in particular. The clock and laser performance assumed here, especially in our most optimistic scenarios, goes beyond what current space-qualified systems can achieve. The requirements this entails, on atom number, the optical link budget, drag-free
performance, and the in-band stability of the clock environment, are detailed
quantitatively in \cref{sec:detector_noise}. Deploying optical lattice clocks in space will require significant advances in miniaturization, power efficiency, and long-term reliability in microgravity. Nevertheless, by showing the capabilities of such a network, our results provide clear motivation for continued technological development. The satellite clock networks analyzed here would also offer a high-precision platform for conducting fundamental physics experiments, enabling novel tests of general relativity and provide sensitivity to dark matter in new regimes \cite{Derevianko2022,Alonso2022,Roberts2017}.

\section*{Acknowledgments}

We thank Evan Hall, Matthew Evans, Jason Arakawa, Jun Ye, Harry Levine and Eugene Knyazev for fruitful discussions and insightful comments on the manuscript. This work was supported by a Packard Fellowship for Science and Engineering, the Sloan Foundation, the Simons Foundation, the Templeton Foundation, the Gordon and Betty Moore Foundation under Grant No. DOI 10.37807/gbmf12966, NASA under Grant No. 80NSSC24K1561, and the National Science Foundation under Grants No. 2143870 and No. 2326810.  D. S. acknowledges support from NSF Grant No. PHY-2020275 [Network for Neutrinos, Nuclear Astrophysics, and Symmetries (N3AS)]. The authors acknowledge the use of large language models in the drafting of this manuscript, in the development and debugging of the numerical analysis code, and in carrying out independent checks of analytic expressions and numerical results. All derivations, results, texts and conclusions were reviewed and verified by the authors.
\appendix

\section{Laser Noise Derivation}

\label{laser_appendix}
In this appendix, we derive the expressions for laser noise that are used in \cref{sec:laser_noise}. For a Ramsey sequence, the filter function is given by,

\begin{equation}
\Hfsq= T^2\sinc^2(\pi fT). 
\end{equation}

Applying this to \cref{eq:sigma_f_general} yields, 

\begin{equation}
   \sigma_f^2 \approx \frac{1}{2T^2}\left(\frac{\Delta_L}{B}+\frac{h\nu}{\eta P} B\right),
   \label{eq:ramsey_laser_noise_B}
\end{equation}
where we have assumed $BT\gg1$ in order to ignore terms of with $e^{-BT}$. We can use this expression to calculate an effective variance in Ramsey phase for a single measurement $\sigma_{\phi}^2 = T^2\sigma_f^2$. Note that this is independent of the Ramsey time.

 We can also extend this treatment of residual laser noise to general pulses sequences. For e.g., a spin-echo sequence, the filter function is obtained by Fourier transforming \cref{eq:spin-echo-window},

\begin{equation}
|H(f)|^2 = \frac{4}{(\pi f)^2} \sin^4\left(\frac{\pi f T}{2}\right)
\end{equation}

We can use the trigonometric identity, $\sin(x)^4 = \frac{1}{8}(3-4\cos(2x)+\cos(4x))$ to evaluate the above integral. Again, ignoring terms with $e^{-BT}$, we get, 

\begin{equation}
   \sigma_f^2 \approx \frac{3}{2T^2}\left(\frac{\Delta_L}{B}+\frac{h\nu}{\eta P} B\right),
   \label{eq:spin_echo_laser_noise}
\end{equation}

Note that this is greater than the noise of a Ramsey sequence by a factor of 3. This could intuitively be understood as noise added from the $\pi$ pulse in the spin-echo sequence. In a Ramsey sequence, we can think of the phase measured in terms of initial and final laser phases $\phi_i$ and $\phi_f$ 

\begin{equation}
  \Delta\phi = \phi_f-\phi_i,
\end{equation}

The condition $BT\gg 1$ ensures that the phases are uncorrelated due to the noise being sufficiently fast compared to the sequence. This lets us assume the same variance $\sigma_\phi^2/2$ for all the phases,  

\begin{equation}
    \sigma_{\Delta\phi}^2 = \sigma_{\phi_i}^2+\sigma_{\phi_f}^2 = \sigma_\phi^2,
\end{equation}

 Now, for the spin-echo sequence, the atoms sample the laser phase $\phi_\pi$ during the $\pi$-pulse. This gives us the measured phase, 
 
\begin{equation}
    \Delta\phi = (\phi_f-\phi_{\pi})-(\phi_{\pi}-\phi_i) = \phi_f+\phi_i-2\phi_\pi.
\end{equation}
Computing the variance of the above expression yields, 
\begin{equation}
    \sigma_{\Delta\phi}^2 = \sigma_{\phi_i}^2+\sigma_{\phi_f}^2+4\sigma_{\phi_\pi}^2 = 3\sigma^2_\phi,
\end{equation}
where we have reproduced the factor of 3 that we obtained using the filter function integral. We can extend this analysis to a general dynamical-decoupling sequence with $N_p$ pulses to get

\begin{equation}
    \sigma^2_{\text{DD}} = (2N_p+1) \sigma^2_\phi
\end{equation}

In the case of 0 dead-time and correlated laser noise between measurements, we can effectively treat a sequence of $n$ spin-echo sequences of length $T$ as one long dynamical decoupling sequence with $N_p = 2n-1$ $\pi$-pulses. The laser noise for this will be given by 

\begin{equation}
 \sigma^2_{n}  = \frac{(4n-1)}{n^2T^2}\sqrt{\frac{h \nu \Delta_L}{P \eta}}  = \left(\frac{4}{\tau T}-\frac{1}{\tau^2}\right) \sqrt{\frac{h \nu \Delta_L}{P \eta}},
\end{equation}

which is generally larger than the noise we expect from uncorrelated measurements (by a factor of 4/3 for large $\tau$). Experimentally, this could be overcome by introducing a small amount of dead-time ($t_D>1/B$).

\begin{figure}
    \centering
    \includegraphics[width=1\linewidth]{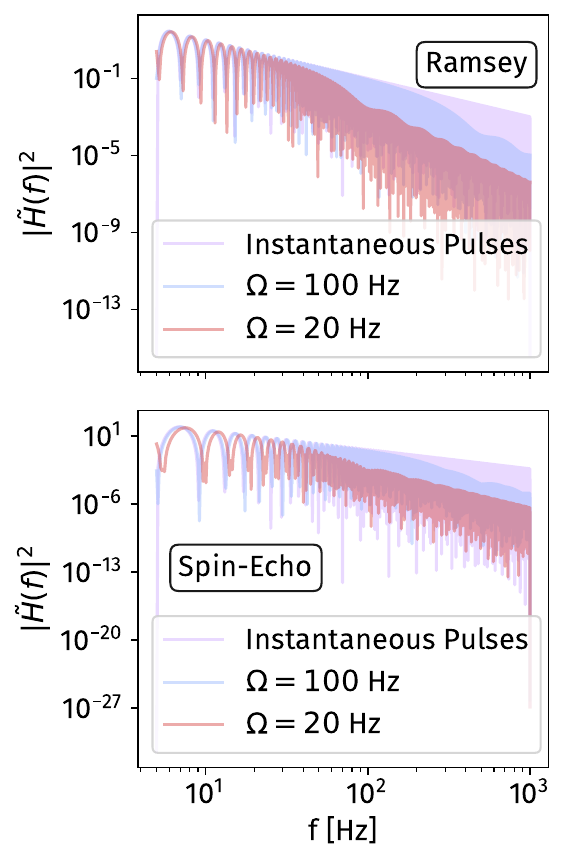}
\caption{Filter function, $\Hfsq$, for 100 second Ramsey (top) and spin-echo (bottom) sequences with finite Rabi pulses with frequency $\Omega$. The finite Rabi frequency modifies the idealized $\text{sinc}^2$ response, with both configurations exhibiting characteristic high-frequency roll-off at $f \gg \Omega/2\pi$.}
    \label{fig:finite_Hf}
\end{figure}

Further laser noise suppression can be achieved by using finite frequency Rabi-pulses. The window function for a Ramsey section with finite non-zero pulse times is given by 

\begin{equation}
    y(t) =
\begin{cases}
\sin(\Omega t) & 0<t <\tau_\frac{\pi}{2}, \\
1  &  \tau_\frac{\pi}{2}<t<\tau_\frac{\pi}{2}+T,\\
\cos(\Omega(t-\tau_\frac{\pi}{2}-T))&  \tau_\frac{\pi}{2}+T<t<2\tau_\frac{\pi}{2}+T,
\end{cases}
\label{eq:ramsey-full-window}
\end{equation}

where $\tau_\frac{\pi}{2} = \frac{\pi}{2\Omega}$ is the pulse time for a $\pi/2$-pulse. The filter function for this sequence is given by,
\begin{equation}
\begin{aligned}
\Hfsq =
T^2
\left[
    \sinc(\pi f T)
    +
    \frac{2}{\Omega T}
    \cos\!\left(
        \pi f T + \frac{\pi^2 f}{\Omega}
    \right)
\right]^2
\\[4pt]
\times
\left[
    \frac{\Omega^2}{
        \Omega^2 - (2\pi f)^2
    }
\right]^2 .
\end{aligned}
\label{eq:finite_rabi_filter}
\end{equation}
The effect of finite Rabi pulses is to modify the filter function through edge terms and a roll-off above the Rabi frequency $\Omega$, shown in \cref{fig:finite_Hf}. At $f = \Omega/2\pi$, the above expression has a limiting value of 

\begin{equation}
\Hfsq_{f\rightarrow\Omega/2\pi} = \left(\frac{\pi\cos(\Omega T/2)+2\sin(\Omega T/2)}{2\Omega}\right)^2
\label{eq:finite_rabi_resonance}
\end{equation}

\begin{figure}
    \centering
    \includegraphics[width=\linewidth]{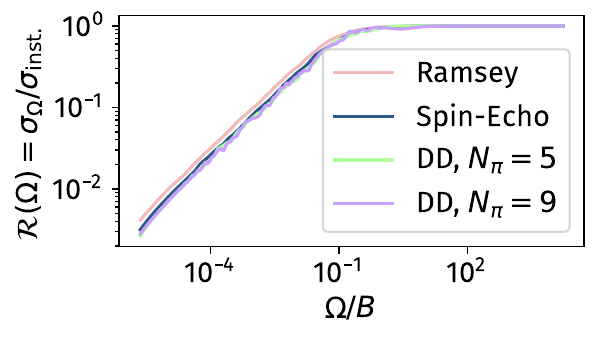}
    \caption{Laser noise suppression from finite Rabi frequency pulses for Ramsey and spin-echo sequences with finite Rabi pulses, normalized to the instantaneous pulse case described in \cref{eq:ramsey_laser_noise_B,eq:spin_echo_laser_noise}. The x-axis shows the Rabi frequency $\Omega$ normalized to the laser noise bandwidth $B$, which is set to 84 kHz. The total sequence time $T$ is 100 s.}
    \label{fig:finite_rabi_noise}
\end{figure}

The filter function noise integral in \cref{eq:sigma_f_general} cannot be evaluated analytically for \cref{eq:finite_rabi_filter}. We can however, numerically evaluate this integral, as is shown in \cref{fig:finite_rabi_noise}, where we introduce a parameter, $\mathcal{R}(\Omega)$ for the noise suppression by finite Rabi pulses. For $\Omega\ll B$, finite Rabi pulses provide noise suppression relative to instantaneous pulses. For large averaging times, $\tau\gg T$, we can use Eq. (C8) of \cite{Kolkowitz16} to obtain an analytic form of the laser noise,

\begin{equation}
	\sigma^2 \approx \frac{1}{(2\pi\nu)^2T\tau}\left(2\frac{h\nu }{P\eta}\Omega\right).
\end{equation}

Comparing this with \cref{eq:laser_noise_alone} for the Ramsey sequence, we see that $\mathcal{R}(\Omega) \approx \Omega/B$.    

This noise suppression is also clear for a spin-echo sequence. We modify the window function in \cref{eq:spin-echo-window} to include finite pulses,

\begin{equation}
y(t) =
\begin{cases} 
\sin(\Omega t), & 0 < t < \tau, \\[1mm]
1, & \tau \le t < \frac{T}{2}+ \tau , \\[1mm]
 \cos\big(\Omega (t - \tau - \frac{T}{2})\big), & \tau + \frac{T}{2} \le t < \frac{T}{2}+2\tau , \\[1mm]
-1, &\frac{T}{2} + 2\tau \le t < T + 2\tau  \\[1mm]
- \cos\big(\Omega (t - 2\tau - T)\big), & 2\tau + T \le t < 3\tau + T, 
\end{cases}
\label{eq:full_spin_echo_window}
\end{equation}

The response function corresponding to this window is given by 

\begin{equation}
\begin{split}
\Hfsq = T^2 \, \frac{16 \, \Omega^2 \, \Big( 
\cos\frac{\pi f (2\pi + T \Omega)}{2 \Omega} 
+ \frac{\Omega}{2\pi f} \sin\frac{\pi f T}{2} 
\Big)^2}%
{((2\pi f)^2 - \Omega^2)^2} \\
\times \, \sin^2\Big( \frac{\pi f T}{2} + \frac{\pi^2 f}{\Omega} \Big)
\end{split}
\end{equation}

In \cref{fig:finite_rabi_noise}, we see that for $\Omega\ll B$, the laser noise for a spin-echo is greatly suppressed compared to the instantaneous pulse case. This suppression pattern extends to higher-order dynamical decoupling sequences as well; the simulated noise for DD sequences with $N_\pi = 5$ and $N_\pi = 9$ pi pulses demonstrates that the $\Omega/B$ scaling holds universally across pulse sequence families. We use our estimates of the laser noise and finite pulse noise suppression to outline a few possible detector scenarios listed in \cref{tab:laser_noise}.

\section{Bayesian Likelihood Construction}
\label{app:bayesian}

In our analysis the simulated detector response data (\cref{eq:complex_response}) are normalized relative to the signal measured by the first detector, leaving us with a relative phase and amplitude response for the entire network. Here, we can also ignore any global phase associated with the network response. 

\begin{equation}
    \tilde{S}_d = S_d / S_0, \quad \tilde{A} = |S_0|.
\end{equation}

We construct a log-likelihood function with three independent Gaussian components

\begin{equation}
    \ln \mathcal{L}(\theta) = \ln \mathcal{L}_\text{S,re} + \ln \mathcal{L}_\text{S,im} + \ln \mathcal{L}_\text{A}
\end{equation}

where the first term is the log-likelihood of the real part of the complex normalized response,
\begin{equation}
    \begin{aligned}
        \ln \mathcal{L}_\text{S,re} = -\frac{1}{2} \sum_{d=1}^{6} \frac{\left[\text{Re}(\tilde{S}_d^\text{data}) - \text{Re}(\tilde{S}_d^\text{model})\right]^2}{\sigma_{\text{re},d}^2} \\
    - \frac{1}{2}\sum_{d=1}^{6} \ln(2\pi\sigma_{\text{re},d}^2),
    \end{aligned} 
\end{equation}

the second term is the log-likelihood of the imaginary part of the complex normalized response,
\begin{equation}
    \begin{aligned}
        \ln \mathcal{L}_\text{S,im} = -\frac{1}{2} \sum_{d=1}^{6} \frac{\left[\text{Im}(\tilde{S}_d^\text{data}) - \text{Im}(\tilde{S}_d^\text{model})\right]^2}{\sigma_{\text{im},d}^2} \\
        - \frac{1}{2}\sum_{d=1}^{6} \ln(2\pi\sigma_{\text{im},d}^2),
    \end{aligned}
\end{equation}

and the third term is the log-likelihood of the overall amplitude,

\begin{equation}
    \ln \mathcal{L}_\text{A} = -\frac{1}{2} \frac{\left(A^\text{data} - A^\text{model}\right)^2}{\sigma_A^2} - \frac{1}{2}\ln(2\pi\sigma_A^2).
\end{equation}

We use nested sampling as implemented in the \texttt{dynesty}~\citep{2020MNRAS.493.3132S} sampler to explore the posterior $p(\hat\theta|\mathbf{d})$ using the the Bayesian inference library \texttt{bilby}~\citep{bilby_paper}. 
Nested sampling simultaneously estimates the posterior and the evidence $\mathcal{Z}$ by evolving a set of $N_\text{live}$ live points through likelihood space, and is particularly well-suited to multi-modal posteriors.

The prior distributions, $\pi(\hat\theta)$ are chosen to reflect the physical parameter space. We use priors that are uniform in right ascension $\alpha \in [0, 2\pi]$ and proportional to $\cos\delta$ in declination ensuring equal probability per solid angle element $d\Omega = \cos\delta\, d\alpha\, d\delta$. The inclination $\iota$ and polarization $\psi$ are assigned uniform priors on the sphere with range $[0, \pi]$ respectively.  The strain amplitude $A$ is assigned a log-uniform prior over $[10^{-21}, 10^{-19}]$. The likelihood, $\mathcal{L}(\mathbf{d} | \theta )$, is a multi-component Gaussian in the frequency domain.

\bibliographystyle{apsrev}
\bibliography{references}


\end{document}